\documentclass[acmsmall]{acmart}

\usepackage{multirow}
\usepackage{graphicx}
\usepackage{enumitem}
\usepackage{caption}
\usepackage{subcaption}
\usepackage{multirow}
\usepackage[linesnumbered,ruled,vlined]{algorithm2e}
\usepackage{xargs}
\usepackage{tikz}
\usepackage{pgfplots}
\pgfplotsset{compat=1.18}
\usetikzlibrary{trees}
\usepackage{mathtools}
\usepackage[colorinlistoftodos,prependcaption,textsize=tiny]{todonotes}
\newcommandx{\unsure}[2][1=]{\todo[linecolor=red,backgroundcolor=red!25,bordercolor=red,#1]{#2}}
\newcommandx{\change}[2][1=]{\todo[linecolor=blue,backgroundcolor=blue!25,bordercolor=blue,#1]{#2}}
\newcommandx{\info}[2][1=]{\todo[linecolor=OliveGreen,backgroundcolor=OliveGreen!25,bordercolor=OliveGreen,#1]{#2}}
\newcommandx{\improvement}[2][1=]{\todo[linecolor=Plum,backgroundcolor=Plum!25,bordercolor=Plum,#1]{#2}}
\newcommandx{\thiswillnotshow}[2][1=]{\todo[disable,#1]{#2}}

\definecolor{ETHc}{RGB}{0,0,0}

\AtBeginDocument{%
  }

\acmJournal{JACM}
\acmVolume{37}
\acmNumber{4}
\acmArticle{111}
\acmMonth{8}

\begin{document}

\title{Efficiently Processing Joins and Grouped Aggregations on GPUs}

\author{Bowen Wu}
\email{bowen.wu@inf.ethz.ch}
\affiliation{%
  \institution{Department of Computer Science, ETH Z{\"u}rich}
  \country{Switzerland}
}

\author{Dimitrios Koutsoukos}
\authornote{Currently at Apple, work reported in this paper was done while the author was at ETH Z{\"u}rich.}
\email{koutsoukos.dimitr@gmail.com}
\affiliation{%
  \institution{Department of Computer Science, ETH Z{\"u}rich}
  \country{Switzerland}
}

\author{Gustavo Alonso}
\email{alonso@inf.ethz.ch}
\affiliation{%
  \institution{Department of Computer Science, ETH Z{\"u}rich}
  \country{Switzerland}
}


\begin{abstract}
  There is a growing interest in leveraging GPUs for tasks beyond ML, especially in database systems. Despite the existing extensive work on GPU-based database operators, several questions are still open. For instance, the performance of almost all operators suffers from random accesses, which can account for up to 75\% of the runtime. In addition, the group-by operator which is widely used in combination with joins, has not been fully explored for GPU acceleration. Furthermore, existing work often uses limited and unrepresentative workloads for evaluation and does not explore the query optimization aspect, i.e., how to choose the most efficient implementation based on the workload. 
In this paper, we revisit the state-of-the-art GPU-based join and group-by implementations. We identify their inefficiencies and propose several optimizations. We introduce GFTR, a novel technique to reduce random accesses, leading to speedups of up to 2.3x. We further optimize existing hash-based and sort-based group-by implementations, achieving significant speedups (19.4x and 1.7x, respectively). We also present a new partition-based group-by algorithm ideal for high group cardinalities. We analyze the optimizations with cost models, allowing us to predict the speedup. Finally, we conduct a performance evaluation to analyze each implementation. We conclude by providing practical heuristics to guide query optimizers in selecting the most efficient implementation for a given workload.
\end{abstract}


\setcopyright{acmlicensed}
\acmJournal{PACMMOD}
\acmYear{2025} \acmVolume{3} \acmNumber{1 (SIGMOD)} \acmArticle{39} \acmMonth{2}\acmDOI{10.1145/3709689}


\maketitle

\section{Introduction}\label{sec:intro}
Nowadays, data centers need to process an ever-increasing amount of data emerging from
 applications such as machine/deep learning (ML/DL),
real-time data analytics, and databases. The increasing demands on computing,
together with the end of Moore's law, have made accelerators very popular~\cite{dally2020domain,Thompson21-decline}. This is true,
especially for GPUs
as they have orders
of magnitude higher parallelism and memory bandwidth than CPUs~\cite{a100,h100,mi250x}, features that
 have attracted interest from database
researchers and developers~\cite{He09-gdb,Lee21-rateupdb,Funke18-pipeline,Yogatama22-orchestrating,Li16-hippogriffdb,
Periklis19-hetexchange,He22-tensor-runtime,Rosenfeld22-revisit,Yuan13-yinyang,Hu22-tcudb,Shanbhag20-crystal,Bress14-cogadb,cudf,Paul16-gpl, Cao-gpudb, kroviakov-crossdevice, boeschen-golap}.
This interest is
reinforced by three trends. First, workloads, such as data pre-processing and ingestion for ML, often comprise many different steps and run on CPUs and GPUs~\cite{Zhao22-dsi}. However, data transfer is one of the
most expensive operations in data centers~\cite{dally2011power,dally2020domain} so it is advantageous to move relational operations to the GPU, where the final results will be processed as part of a downstream application, e.g., training an ML model. Second,
the increased hardware capabilities of GPUs could potentially be used to accelerate costly relational
operations like joins. 
And third, GPUs are now pervasive, with growing support in terms of development and tools. It makes sense to take advantage of these investments in hardware and software for relational operators as they are being done around, e.g., tensor computations \cite{Supun20-hummingbird,tensorflow2015-whitepaper,pytorch}.

\begin{figure}[t]
  \centering
  \includegraphics[width=.6\linewidth]{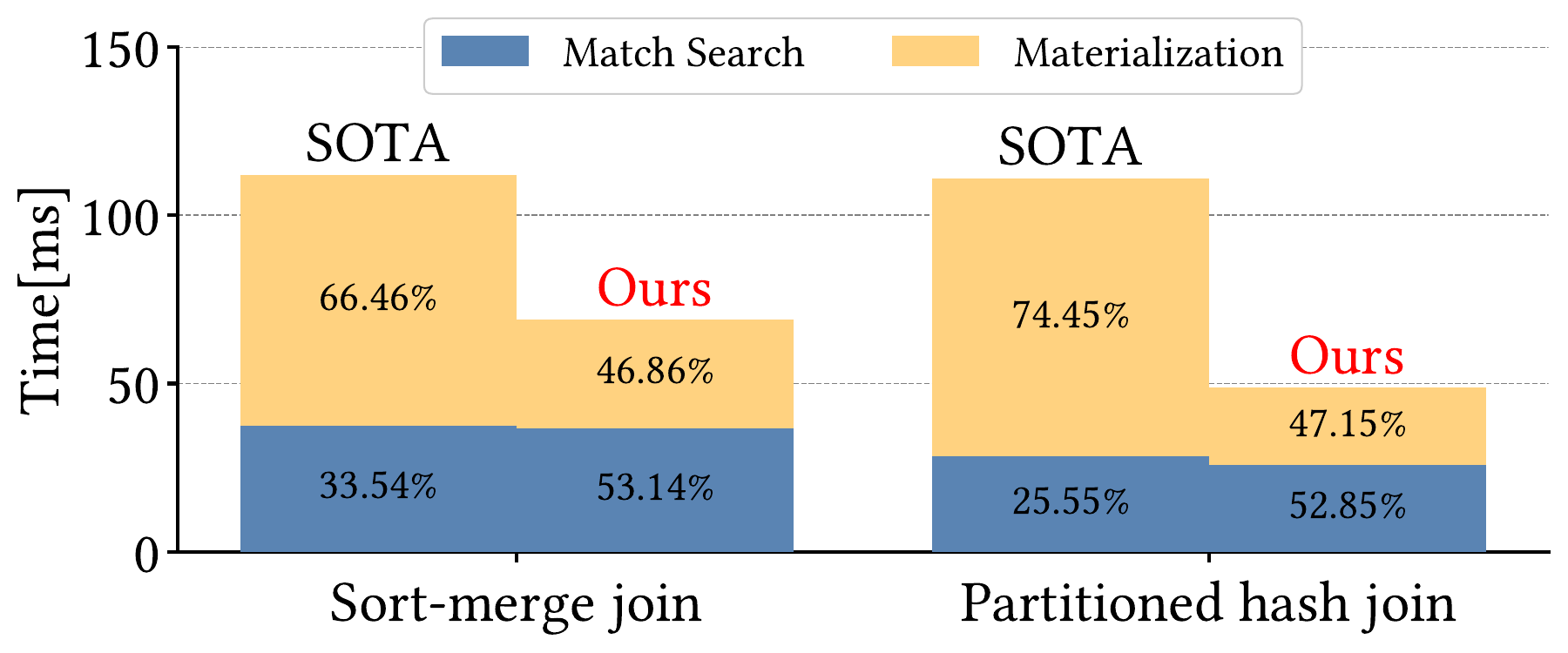}
  \caption{Time break-down for GPU-based join processing.}
  \label{fig:intro}
\end{figure}
These factors have triggered recent efforts to optimize
relational operators in GPUs, e.g., joins~\cite{sioulas19-partitioned-radix-join,Paul20-revisit-gpujoin,Yabuta17-gpujoin,Rui17-fastequijoin} or grouped aggregations~\cite{Diego18-groupby, Tomas15-groupby}.  
To better understand the role of GPUs in relational data processing, in this paper, we revisit GPU-based join and group-by implementations and identify four major areas of improvement.

\textbf{(1)} We observe that random global memory accesses contribute the most to the execution time of all existing algorithms. Random accesses are common in an often overlooked stage of an algorithm -- writing the output relation into the global memory. On the other hand, other stages, e.g., finding matching tuples or groups, traditionally thought to be harder, are well-optimized, making output production an even more serious bottleneck. For instance, we analyzed two recent~\cite{sioulas19-partitioned-radix-join,Rui17-fastequijoin} hardware-conscious
join implementations on GPUs (Figure~\ref{fig:intro}) by letting them join a primary-key relation of 1.5 GB with a foreign-key relation of 3 GB on an A100 GPU, each having two columns of payloads and then materialize the results.
Due to random accesses, the cost of materialization is very significant and can contribute up to 75\% of the overall runtime.
{\color{ETHc} Previous work overlooks the materialization costs in the GPU probably because some assume a CPU-GPU hybrid processing mode where the materialization can happen in the CPU. Alternatively, others study processing data beyond the capacity of GPU memory, in which case other costs, e.g., data transfer, are more dominant. Our work focuses on the case where the operators are only executed by the GPU and in the GPU memory.}

\textbf{(2)} A high proportion of TPC-H/DS queries have a group-by as the next step after a sequence of joins. It is, thus, beneficial to process both in the GPU to avoid extra data movement between operators. However, not enough work has been devoted to aggregation on GPUs. An important aspect is whether the optimizations done for the join (e.g., partitioning) can also be applied to group-by. However, despite its similarity with the join, the grouped aggregation (group-by henceforth) poses unique challenges requiring customized optimizations. 

\textbf{(3)} Existing algorithms are evaluated with a rather narrow spectrum of workloads. The limited workload selection not only masks potential inefficiencies but also leads to biased optimizations. For example, GPU-based joins are commonly evaluated with input relations that have only one key and one non-key, both being 4-byte long. In this scenario, searching for matching tuples is the only dominant cost and sort-merge joins and partitioned hash joins perform well. However, as shown in Figure~\ref{fig:intro}, when evaluated with more non-key columns, the materialization cost of such implementations soars because their optimized match searching introduces more random accesses during materialization. In general, different data types, data distribution, and the number of non-key columns can have a big impact on the performance. Their effects need to be accounted for in order to design better algorithms. 

\textbf{(4)} As an important function of the query optimizer, deciding the most efficient implementation of an operator for a given query plan and workload is missing. Existing work does not compare alternative, well-optimized implementations of each operator and therefore provides limited insight on when to use one algorithm over another.

This paper addresses (1)-(4) with the following contributions:

\begin{itemize}[wide, labelwidth=!, labelindent=0pt, topsep=0pt]
    \item We identify the main source of inefficiency in existing GPU-based join and group-by implementations to be the random accesses of \emph{non-key} columns. In light of this, we propose an optimization technique, applicable to sort-/partition-based implementations, that substantially reduces random accesses and improves the overall performance by up to 2.3x in joins and 2.2x in group-by.
    \item For partition-based join and group-by implementations, we propose to substitute the bucket-chain-based partitioning from previous work ~\cite{sioulas19-partitioned-radix-join} with \emph{radix-sort-based} partitioning that provides not only the same performance but also stability\footnote{Stability as in stable sorting algorithms.} and constant-time lookup-by-index into the partitioned data, both of which are indispensable for supporting the optimizations proposed in this paper. Moreover, radix-sort-based partitioning is immune to data skew, unlike previous work. 
    \item We present optimization techniques for the group-by, including, shared-memory-based aggregation, partition-based group-by, and dictionary encoding. These techniques improve the state-of-the-art by up to 19.4x, 2.88x, and 2.63x, respectively. 
    \item We provide cost-based models to understand the performance gain and characterize the workloads for which they are suitable. Our experimental evaluation backs the accuracy of the models.
    \item Using a diverse collection of workloads, we comprehensively evaluate the performance of the state-of-the-art join and group-by implementations and our own optimized ones. We summarize our findings with heuristics, e.g., decision trees, that can guide query optimizers in choosing the best algorithm for a specific workload.
\end{itemize} 
\section{Background}\label{sec:background}
\subsection{GPU Architecture and CUDA}\footnote{We focus on NVIDIA GPUs. Other GPUs, e.g., AMD, work similarly.}
A GPU consists of multiple streaming
multiprocessors (SM). 
An SM groups every 32 threads into a warp. Threads in a warp work in a single-instruction-multiple-threads (SIMT) manner. A \emph{thread block} can contain multiple warps and is always scheduled to run on a single SM. A CUDA kernel normally contains many thread blocks to fully utilize the GPU resources.

The GPU memory hierarchy consists of register files (exclusive to a thread), L1 cache (exclusive to an SM), L2 cache (shared by all SMs), and global memory. 
In each SM, part of the L1 cache can be configured as \emph{shared memory}, shareable within a thread block. Developers can programmatically cache frequently used data in the shared memory to reduce global memory accesses. Efficient memory accesses are crucial for performance. \emph{Coalesced} memory accesses, (i.e., spatial locality within a warp) allow the GPU to combine loads/stores from the same warp into as few cache-line accesses as possible. Uncoalesced accesses, e.g., random accesses, degrade performance significantly.

\subsection{In-memory Join and Group-by on GPUs}
\label{sec:background-join}
In this paper, we assume that both the input and output of the operators fit into the device memory. 
For the join, we focus on the most common inner equal join. It takes two input relations $R(key, r_1, \dots)$
and $S(key, s_1, \dots)$, enumerates pairs of tuples sharing the same key and produces an output relation, $T(key, r_1, \dots, s_1, \dots)$. $R.key$ and $S.key$ are the join key, and $r_1, \dots, s_1,
\dots$ are called \emph{payload} columns or non-key attributes. For simplicity, we further assume that $R.key$ is a primary key and $S.key$ is a foreign key. We will mainly focus on primary-foreign-key join for the rest of the paper, but we also discuss the many-to-many join in the evaluation section. For the GPU, the join can be processed with non-partitioned hash join, partitioned hash join, and sort-merge join. 

The group-by deals with one input $R(key, r_1, \dots, r_n)$ with $key$ being the group key. We assume there is only a single group key since multiple group keys can be normalized and combined into a single one, a common approach in databases~\cite{velox,key-normalization}. We denote the number of groups that $R$ has as $g$, which is \emph{unknown} to us. In this work, we assume that the aggregation functions are all reducible, e.g., sum, max, and min. Existing GPU-based group-by operators are implemented with hash-based or sort-based algorithms. 

\subsection{GPU Primitives}
\label{sec:gpu-primitives}

GPU primitives are
procedures frequently present across many applications. These primitives, developed by GPU manufacturers~\cite{cub,nvidia/thrust},
provide high-performance out-of-the-box and are tuned automatically for
different GPU architectures.
The primitives used in this paper are the following:

\begin{itemize}[wide, labelwidth=!, labelindent=0pt, topsep=0pt]
  \item \textsc{\textbf{radix-partition}}$(kin,vin,kout,vout,i,j)$. Partitions the key ($kin$) and value ($vin$) arrays, based on the radix bits of keys (i.e., the $i$-th to $(j-1)$-th bit in the binary representation). It stores the results in $kout$ and $vout$ arrays. In the result arrays, partitions are \emph{consecutively} stored, making $kout$ ($vout$) the same size as $kin$ ($vin$).
      \textsc{radix-partition} is stable, meaning that identical keys $k_1$ and $k_2$ in $kin$ will maintain their relative order in $kout$.  

      CUB~\cite{cub} implements this primitive in the following way. It starts with building a histogram on keys, which requires one sequential pass of $kin$. Then, it performs the partitioning in multiple passes. For Ampere architecture, a maximum of 256 partitions
      (i.e., using max. 8 radix bits) can be produced each pass. Therefore, $\lceil (j-i)/8\rceil$ passes are needed. 
      For each pass, it reads in $kin$ and $vin$ once and writes out $kout$ and $vout$ once, all in a \emph{sequential and coalesced} way. This allows each pass to utilize the memory bandwidth very efficiently.
      
      Note that when $i=0$ and $j=\text{sizeof}(key)\times 8$, the primitive becomes \emph{radix sorting}.

  \item \textsc{\textbf{gather}}$(in,map,out)$. For all valid $i$'s, this primitive computes:
      $out[i]\gets in[map[i]]$. 
      Reading $in$ may cause random memory accesses if the gather map $map$ lacks ``ordering''~\cite{He07-multipassgather}. 
      Depending on the efficiency of reading $in$, a
      \emph{clustered} gather exhibits a low cache miss rate and highly coalesced memory accesses, compared to an
      \emph{unclustered} gather that exhibits the opposite. Unclustered \textsc{gather}s often appear in the materialization or aggregation of existing join and group-by implementations. 
\end{itemize}
\section{Existing GPU-based Join and Group-By}\label{sec:impl}

In this section, we describe in detail how the most efficient GPU-based join and group-by algorithms work. By analyzing their implementation details, we illustrate their inefficiencies and their causes. 
Besides the assumptions mentioned in Section~\ref{sec:background-join}, we also assume that relations are stored
in the GPU device memory as columns, and all columns are stored as arrays. 

\subsection{Join Implementations}\label{sec:join-impl}

We analyze existing sort-merge joins, partitioned hash joins and non-partitioned hash joins on the GPU. We focus on ``wide'' joins where both $R$ and $S$ have more than one non-key attribute. Wide joins are the most general case and entail the materialization of non-key attributes (aka payloads) of \emph{both} input relations. Previous work primarily focuses on ``narrow'' joins where $R$ and $S$ have only one payload column each. The materialization of narrow join can be inlined with the match search and is therefore inexpensive.

\subsubsection{Sort-Merge Join}\label{sec:unopt-smj}
The sort-merge join contains three phases. In the first phase (transformation phase), it first materializes the
\emph{physical} tuple identifiers $ID_R$ and $ID_S$ for $R$ and $S$ and then sorts
the pairs of columns, $(R.key, ID_R)$ and $(S.key, ID_S)$, to produce
intermediate relations $R'$ and $S'$, respectively. During sorting, it can use
the \textsc{radix-partition} primitive to perform a radix sort.

The second phase (match-finding phase) identifies matching tuples between the sorted $R'$ and $S'$. This phase produces the intermediate relation $T'(key, ID_R, ID_S)$, which contains keys and
IDs of matching tuples. To ``merge'' $R'$ and $S'$ efficiently on the GPU, Rui et
al.~\cite{Rui17-fastequijoin} and ModernGPU~\cite{moderngpu} use the Merge Path algorithm~\cite{Green12-mergepath}. This algorithm divides both
$R'$ and $S'$ into non-overlapping partitions, where each partition pair
is assigned to a GPU thread. 
The algorithm ensures that the partition pairs can be merged independently.
It also guarantees that threads get approximately the same amount of work
regardless of data distribution, which makes this algorithm resilient to data
skewness. Both Rui et al. and ModernGPU apply the Merge Path algorithm twice: once to find the
lower bound of each key of $S'$ in $R'$, and a second time
to determine the upper bound. The matches for a key in $S'$ fall
within the range defined by these two bounds. However, in a primary-foreign-key join, a foreign key can have at most one matching primary key. In such a case,
we only need to apply the Merge Path algorithm once to find either upper or
lower bounds.

In the last materialization phase, the $ID_R$ and $ID_S$ from $T'$ are used to
fetch matching tuple payloads from $R$ and $S$ and copy these values into the
corresponding columns of the final output $T$. The process is repeated for all
payload columns in both input relations using the \textsc{gather} primitive.
However, $ID_R$ and $ID_S$ represent tuple IDs from \emph{unsorted} relations.
They become
randomly permuted during sorting. As a result, the
\textsc{gather} operations become \emph{unclustered},
leading to a significant increase in uncoalesced memory accesses and cache misses.

\subsubsection{Partitioned Hash Join}\label{sec:unopt-phj}
\begin{figure}[t]
    \centering
    \includegraphics[width=.75\linewidth]{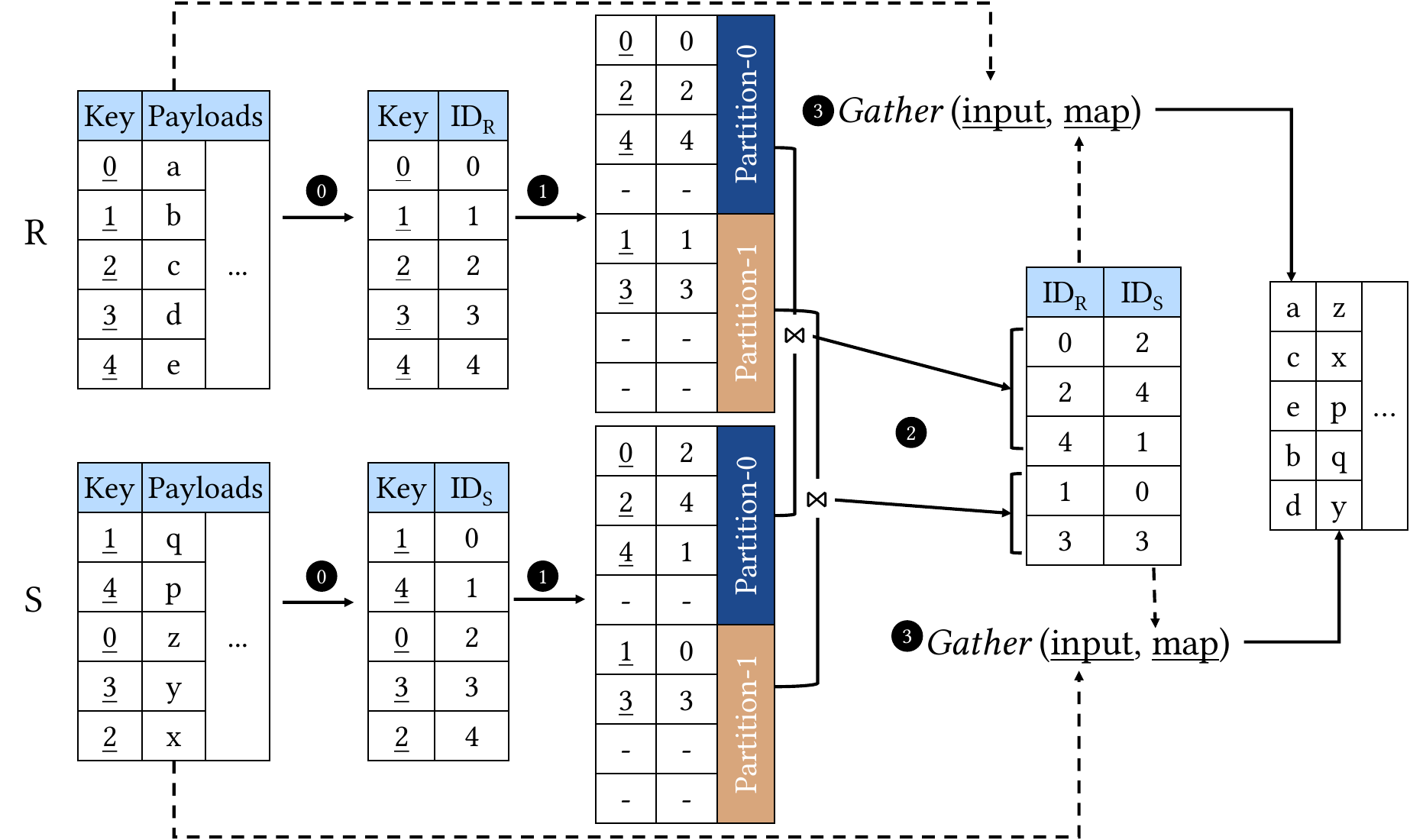}
    \caption{Partitioned hash join proposed by Sioulas et al.}
    \label{fig:phj-unopt}
\end{figure}

The partitioned hash join contains the same three phases as the sort-merge join. In the transformation phase, the partitioned hash join algorithm begins with
materializing the \emph{physical} tuple identifiers, $ID_R$ and $ID_S$, for the relations
$R$ and $S$. It then proceeds to partition pairs of columns,
namely $(R.key, ID_R)$ and $(S.key, ID_S)$, resulting in the intermediate
relations $R'$ and $S'$. Sioulas et al.~\cite{sioulas19-partitioned-radix-join} introduced a multi-pass radix
partitioning implementation (Figure~\ref{fig:phj-unopt}), which
utilizes bucket chains to store and manage partitions. In this implementation,
a bucket is a fixed-size, pre-allocated memory region, and it
exclusively belongs to a single partition. Buckets of the same partition
are connected through a chain, and each bucket maintains the location
information of the next bucket in the chain. During the partition, the algorithm allocates an initial bucket
for each partition. If a partition's size exceeds the capacity of a single
bucket, new buckets are dynamically allocated and added to the chain. Although
all buckets are allocated by ``offsetting'' a pre-allocated buffer, empty memory
space exists between buckets since not all buckets are filled up.
For example, in Figure~\ref{fig:phj-unopt} the partition 0 cannot fill the
entire bucket, and therefore, a gap exists between partition 0 and 1, leading to memory \emph{fragmentation}. While partitioning, each thread block constructs a histogram in the shared memory for its work set using the radix bits. Subsequently, all thread blocks use atomic operations to determine the partition sizes and the locations to write the partitions. The partitioning requires multiple passes because large relations
need a high fan-out to keep each partition sufficiently small such that it fits
in the shared memory.

Partitioning with bucket chaining performs well because it leverages the shared
memory and uses atomic operations to determine where each key
should be written. Despite its high performance, the use of atomic operations
can result in \emph{non-deterministic} behavior~\cite{adinets2022onesweep}, i.e., partitioning produces 
different outcomes across different runs. This occurs because the algorithm
does not calculate prefix sums to decide where to write each tuple inside a partition;
instead, a tuple is written when the thread has finished processing it.

Before the match-finding phase, Sioulas et al. balance the workloads among
thread blocks. They decompose large probe-side partitions with many buckets
into sub-partitions of fewer buckets. Then, they proceed to join
co-partitions as follows: a thread block reads a bucket from the build-side
partition builds a hash table with it in the shared memory, and then probes
the hash table with keys streaming from the probe-side partition.
If a build-side partition has more than one bucket, they repeat the
aforementioned procedure for each build-side bucket, in an operation resembling
a block nested loop join. The hash join is highly efficient because the
hash tables reside in the shared memory, and that reduces
random accesses into the global memory.

The materialization process resembles that of sort-merge join. 
The \textsc{gather} is still unclustered, since the tuple identifiers, $ID_R$
and $ID_S$, correspond to tuples in the original, non-partitioned input
relations, which have been permuted during partitioning. 

\subsubsection{Non-partitioned Hash Join}\label{sec:nphj}
The non-partitioned hash join only contains two phases. In the match-finding phase, the algorithm begins with materializing the tuple IDs $ID_R$ and $ID_S$ for relation R and S, respectively. Then, it inserts $R.key$ and $ID_R$ into a global hash table. After that, it checks for each $S.key$ if there is a match by probing the hash table. The result of this step is an intermediate relation $T'(key, ID_R, ID_S)$. In the materialization phase, it again uses the \textsc{gather} primitive to collect non-key attributes of $R$ and $S$ into the final output with $T'.ID_R$ and $T'.ID_S$ respectively. Compared to the materialization phase of the sort-merge join and partitioned hash join, random accesses only happen when materializing $R$'s payload data but not for $S$. 

\subsection{Group-by Implementations}\label{sec:groupby-impl}
The GPU-based group-by has received less attention than joins. 
Nevertheless, it presents some unique challenges: (1) its performance is highly sensitive to the number of groups~\cite{Tomas15-groupby,rosenfeld-hash-groupby}; (2) many aggregation functions (e.g., sum and max) mandate accessing all values of a payload column, potentially leading to excessive random accesses that joins with a low match ratio will not face.

\subsubsection{Hash-based Group-by}\label{sec:hash-group-by}
The hash-based group-by uses a global hash table to keep track of unique group keys and their corresponding aggregates.
This approach is adopted by cuDF for cases where the aggregation can be computed using atomic operations.  
Instead of storing aggregates directly in the hash table\footnote{GPU-based hash tables usually cannot store values larger than 8 bytes due to restricted atomic operations.~\cite{cuCollections}}, cuDF stores them in a separate array $arr$ (one array per aggregation), which has as many entries as the input. If $k=key[i]$ is inserted for the first time, then the hash table will save the pair $(k,i)$ so that subsequent insertions of key $k$ will read position $i$ and aggregate the payload data into $arr[i]$. After insertions are done, cuDF will extract all pairs of $(key, pos)$ from the hash table and use the positions to \emph{gather} the aggregates from the arrays. 

The hash-based group-by from cuDF has two main inefficiencies. First, both inserting values into a global hash table and aggregating values in the arrays introduce random and uncoalesced accesses into the global memory, causing a low memory bandwidth utilization. This can be alleviated when the number of groups is low since non-empty hash table entries can be cached in the L2 cache. Moreover, with a low number of groups, the work set of a warp will contain fewer different keys reducing the number of uncoalesced memory accesses. However, when the number of groups is very small, hash table operations as well as aggregating values will suffer from severe contentions due to atomic operations. This conflicting effect of group cardinality is also observed in~\cite{Tomas15-groupby}. Second, gathering aggregates in the last step also exhibits poor memory access patterns and will cause significant performance penalties when the number of groups is high. In summary, the hash-based group-by works the best when the number of groups strikes a balance between contention and poor memory accesses.

\subsubsection{Sort-based Group-by}\label{sort-groupby-gfur}
The sort-based group-by has been described by He et al. in ~\cite{He22-tensor-runtime} and has also been implemented in cuDF for cases where hash-based group-by cannot be used. He et al. first sort the group keys with tuple IDs and then use the IDs to gather the payload data. After sorting the group keys, they identify unique keys, retrieve their payload data based on tuple IDs, and aggregate them. cuDF uses a similar implementation. The greatest inefficiency in such an implementation is to use the tuple IDs to gather the payload data which will cause excessive random and uncoalesced memory accesses for a high number of groups.

\section{Optimization Techniques}
\label{sec:opt}
In this section, we present optimizations to the existing GPU-based join and group-by implementations. 
We first discuss how to reduce random accesses that appear in all existing implementations. Then, we propose several optimizations targeting the group-by. For workload-dependent optimizations, we also present cost models to quantify the performance improvement and characterize workloads for which these optimizations are suitable. 

\subsection{Tackling Random Accesses}
\subsubsection{Causes of Random Accesses}\label{sec:random-access-cause}
All the algorithms analyzed in the previous section contain random global memory accesses that grow proportionally with the input or output relation size.
There are two main causes of this inefficient access. The first arises from manipulating a global hash table (e.g., insertions and lookups). Sort-based and partition-based implementations avoid it by sorting or partitioning the join key or group key so that no global hash table is needed. For generalization purposes, in this context, we call both sorting and partitioning \emph{transformations}, since they change the order of the original keys. Transformation is the second cause of random access. To be more specific, while the transformation makes the match searching and group discovery more efficient by avoiding using a global hash table, it results in more random accesses into the \emph{untransformed} payload data during the materialization and aggregation. For example, during the
materialization phase, both the sort-merge join and partitioned hash join use inefficient \emph{unclustered} \textsc{gather}
procedures, because tuple IDs are randomly permuted as they transform the keys. In general, all existing sort-/partition-based algorithms exhibit the workflow shown in Figure~\ref{fig:gfur}, which we call ``gather-from-untransformed-relations'' (GFUR). GFUR is inefficient because of the use of permuted tuple IDs to access the payload data ($p[ptid[i]]$). The optimization we propose targets random accesses in GFUR to make sort-/partition-based implementations \emph{completely} free of random accesses. 

Before diving into the optimization, we first explore how expensive the unclustered \textsc{gather} is. We use NVIDIA Nsight compute profiler to break down the performance. Table~\ref{tab:gather-profile} shows the performance of gathering $2^{27}$ 4-byte elements on an A100 GPU. Although having the same number of instructions, the clustered \textsc{gather} is 8.52x faster than the unclustered one. Suppose the peak global memory bandwidth is $B_G$ and randomly accessing global memory has bandwidth $B_R$. Since both clustered and unclustered \textsc{gather}s read the gather map and write the output sequentially, they share the common costs of $(2\times 4\text{ bytes} \times 2^{27}/B_G$). To read the input, the clustered and unclustered \textsc{gather}s spend $(4\text{ bytes} \times 2^{27}/B_G)$ and $(4\text{ bytes} \times 2^{27}/B_R)$, respectively. Given that the clustered one is 8.52x faster, we get $B_G \approx 24 B_R$.

\begin{table}[h]
    \centering
    \caption{Unclustered vs clustered \textsc{gather}s.}
    \label{tab:gather-profile}
    \resizebox{0.7\linewidth}{!}{%
    \begin{tabular}{@{}lll@{}}
    \toprule
     & Unclustered \textsc{gather} & Clustered \textsc{gather} \\ \midrule
    Number of items                    & $2^{27}$   & $2^{27}$   \\
    Total cycles                       & 12,052,942 & 1,414,398  \\
    Number of warp instructions        & 77,594,624 & 77,594,624 \\
    Avg. cycles per warp instruction   & 1037.06    & 115.74     \\
    \bottomrule
    \end{tabular}%
    }
\end{table}

\begin{figure}
    \centering
    \begin{subfigure}[t]{0.48\linewidth}
    \centering
    \resizebox{\linewidth}{!} {
        \begin{tikzpicture}
            \node (A) {\shortstack{Transform (keys $k$, tuple IDs $tid$) to get\\(transformed keys $tk$, permuted tuple IDs $ptid$)}};
            \node[below=of A] (B) {For valid $i$-s, access $tk[i]$ and its payload data $p[ptid[i]]$};
            
            \draw[->] (A) -- (B) node[midway, right] {};
        \end{tikzpicture}
    }
    \caption{\textit{GFUR} pattern.}
    \label{fig:gfur}
    \end{subfigure}
    \hfill
    \begin{subfigure}[t]{0.48\linewidth}
    \centering
    \resizebox{\linewidth}{!} {
        \begin{tikzpicture}
            \node (A) {\shortstack{Transform (keys $k$, payload $p$) to get\\(transformed keys $tk$, transformed payload $tp$)}};
            \node[below=of A] (B) {For valid $i$-s, access $tk[i]$ and its payload data $tp[i]$};
            
            \draw[->] (A) -- (B) node[midway, right] {};
        \end{tikzpicture}
    }
    \caption{\textit{GFTR} pattern.}
    \label{fig:gftr}
    \end{subfigure}
\end{figure}

\subsubsection{Removing Random Accesses}
Since the targeted random accesses are mainly caused by unclustered \textsc{gather}s, we need to make the \textsc{gather}s more clustered. To achieve this, we propose a different workflow than GFUR, called ``gather-from-transformed-relations'' (GFTR) shown in Figure~\ref{fig:gftr}. In GFTR, we transform the payload data together with the keys (join key or group key) using the \emph{same} transformation procedure. It is worth emphasizing that we use the actual sorting or partitioning procedure to transform the payload data instead of transforming the (keys, tuple IDs) once and permuting the payload data using \textsc{gather}s, which is exactly GFUR.
{\color{ETHc} The core observation and idea behind GFTR resemble those behind hardware-conscious join algorithms~\cite{Balkesen13-partitioned-hash-join, sioulas19-partitioned-radix-join, Shatdal94-phj, Manegold02-phj}, where more efficient hash table operations can pay off the additional cost of partitioning the relations. In GFTR, we perform extra work to transform payloads in order to make the materialization free of inefficient random accesses, which are the bottleneck of the system.}

GFTR removes random accesses from sort-/partition-based implementations as the 
transformation itself contains only sequential accesses (see Section~\ref{sec:gpu-primitives}). In addition, transformed payload data will be accessed sequentially as long as the transformed keys are processed sequentially. 

While the GFTR avoids the random accesses, it brings another cost -- sorting or partitioning \emph{all} payload data. GFTR pays off when transforming all payload data is cheaper than randomly accessing payload data. In the following analysis, we will show that this depends on the workload and can be understood by modeling. 

Next, we will use the sort-based group-by and partition-based join to illustrate the idea of GFTR, discuss critical implementation details, and analyze how much performance gain/loss it can bring.  

\subsubsection{Application to Sort-Based Group-by}\label{sort-groupby-gftr}
Applying the GFTR technique to the sort-based group-by only needs to additionally \emph{sort} the payload data according to the group key. After sorting, the payload values from the same group are clustered together, and therefore aggregating them accesses the memory sequentially.

We can use a cost model to understand when using GFTR is better. Intuitively, the GFTR version works better for a large number of groups since the payload data will be permuted in a more random way during the gathering of GFUR. For a small number of groups, the payload data will be permuted less aggressively, resulting in a better memory access pattern. Suppose the input relation contains $N$ tuples, distributed evenly across $g$ groups. We further assume that group keys and the payload data are all 4-byte long. The peak memory bandwidth is $B_G$ bytes/second. To sort the 4-byte payload data with respect to keys, we need to scan the keys once to build histograms and then perform radix partitions per 8 bits (4 passes in total). In each pass, we need to read the unsorted keys and values and write out the sorted ones. Therefore, in total, sorting one pair of keys and payload requires 17 reads and writes of keys or payload data. Each pass of keys or payload data is sequential and therefore can be assumed to achieve the peak memory bandwidth. The cost of sorting ($C_S)$ is hence $C_S = (17\times N\times 4)/B_G= 68N/B_G$. 

To model the performance of accessing payloads with permuted tuple IDs (essentially an \emph{unclustered} \textsc{gather}), we calculate how much the peak memory bandwidth $B_G$ is discounted due to non-sequential accesses. Consider the work set of a warp containing 32 permuted tuple IDs used to read 32 unsorted payload items. Given that the input contains $g$ groups, the expected number of unique groups in a warp's work set can be approximated by $u=\min (32, g)$. We further assume that only the payload items from the same group will reside in the same cache line. Then we know that to read 32 payload items, the warp needs to read $u$ cache lines leading to an approximate memory bandwidth of $(B_G/u)$ bytes/second. In Section~\ref{sec:random-access-cause}, we show that the random memory access bandwidth is around $B_G/24$. Together with the cost of writing payload data to memory and reading tuple IDs (both sequentially), the cost of \emph{unclustered} \textsc{gather} ($C_P$) is:
\begin{equation*}
\begin{split}
\underbrace{\frac{2N\times 4}{B_G}}_{\text{Read IDs and write results}} + \underbrace{\frac{N\times 4}{\max(B_G/u,B_G/24)}}_{\text{Read payload (permuted)}} = \frac{(4\min(24, g)+8)N}{B_G}.
\end{split}
\end{equation*}

The model implies that we need at least 16 groups to make the sorting faster than the permutation.

\subsubsection{Application to Partitioned Hash Join}\label{sec:opt-phj}
Given that the adaptation of the sort-based group-by is straightforward, it is
tempting to likewise adapt the existing partitioned hash join by partitioning
all payload columns together with their keys.
However, this optimization fails because of two properties of bucket-chain
partitioning mentioned in Section~\ref{sec:unopt-phj}, namely non-determinism
and fragmentation. Firstly, during partitioning, a thread block uses
\emph{atomic} operations to allocate new buckets and determine where to write in the buckets.
Atomic operations do not produce identical results
across runs. For example, consider thread A and B both calling \textsc{atomicAdd}
on the \textsc{offset} variable to determine at which positions they should
write the output values. It is impossible to predict whether thread A's output
value appears before or after B's because the order depends on which thread
returns from the \textsc{atomicAdd} first. Therefore, partitioning the pair
$(key,col_1)$ might have different results than
partitioning $(key,col_2)$ even though both are valid partitions.
Non-determinism can lead to wrong join results. 
Additionally, even
if we make the partitioning deterministic, the representation of partitions as
chains of buckets makes it impossible to quickly look up values by indices
in a partitioned column due to fragmentation. To access the $i$-th value in the
partitioned payload column, we first need to calculate to which partition and
which bucket it belongs and the offset into that bucket. All these additional
calculations make the materialization inefficient.

\begin{figure}[t]
    \centering
    \includegraphics[width=.75\linewidth]{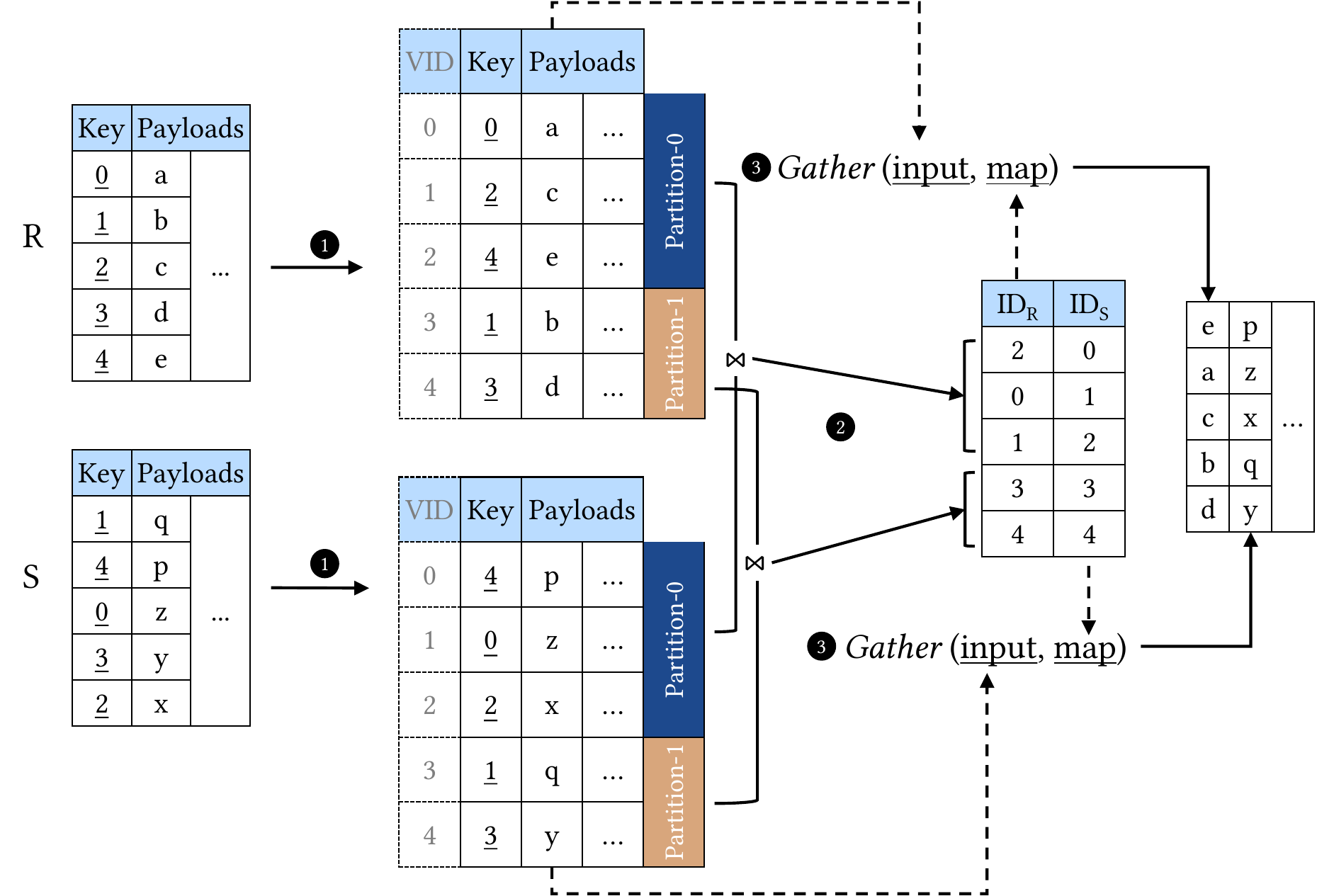}%
    \captionsetup{belowskip=0pt}
    \caption{Optimized partitioned hash join.}%
    \label{fig:phj-opt}
\end{figure}

To apply the GFTR pattern in the partitioned hash join, we need to ensure that
the partitioning produces consistent results across $(key,col_1),\dots,(key,col_n)$
and to look up values of a partitioned column by indices in constant time. To
this end, we propose a new partitioned hash join implementation that overcomes
these difficulties (Figure~\ref{fig:phj-opt}).
The partitioning phase (Step 1) uses the \textsc{radix-partition} primitive to
partition the input relations $R$ and $S$ into \emph{contiguous} arrays, where values from the
same partition are grouped together in the transformed relations $R'$ and $S'$.
Because the radix sort requires the \textsc{radix-partition} to be
stable~\cite{adinets2022onesweep}
, it will
always produce the same partitioning results given the key column. That means that if $k_1$ occurs before $k_2$ in the input and they belong to the same partition, then $k_1$ should also occur before $k_2$ in the result.

Nonetheless, the boundaries between partitions are unknown after the radix
partitioning. Thus, we need additional calculations to figure out the offset
and size of each partition and we do so by constructing a histogram to find the size 
of each partition. Now that we know the partition size, we can determine the offset of each
partition by calculating a prefix sum. To prevent partitions from being large
and causing unbalanced workload distribution during the hash join, we
re-partition them into smaller sub-partitions.
Afterwards, in step 2 (Figure~\ref{fig:phj-opt}), we find the matches between the keys of $R'$ and $S'$, after which identifiers of matched tuples are
produced. Because the partitioned column
is now an array instead of fragmented buckets, looking up by indices can be done in constant time, and
\textsc{gather}s become efficient (Step 3).
The additional transformation cost of this optimized partitioned hash join is partitioning all non-key attributes.
For partitioning, we use 15 or 16 radix bits to determine a partition, resulting in two invocations of \textsc{radix-partition} per payload column. Compared to four invocations of \textsc{radix-partition} in sorting, partitioning is more efficient.

Theoretically, the optimized partitioned
hash join guarantees that only the matching tuple identifiers of the probe
relation are highly clustered, whereas those of the build relation are not as
clustered. In practice, gathering the payloads from the build
relation is also greatly improved compared to the unoptimized version.

It is important to clarify that while we propose a new partitioned hash join 
implementation to support the GFTR pattern, the implementation itself is not 
limited to GFTR. This implementation is flexible enough to follow the
GFUR pattern as well by not partitioning the payload columns. This flexibility
makes it competitive for workloads where the GFUR pattern performs better,
such as joins with a low match ratio.

Next, we will use a model again to understand when GFTR is better than GFUR. Instead of focusing on the effect of groups as in Section~\ref{sort-groupby-gftr}, we analyze the effect of the match ratio of the join. We compare two cases -- (1) gather M out of N different elements of a column in an unclustered way and (2) partition the same column with its keys first and then gather M out of N different elements of the partitioned column in a clustered way. Both cases simulate materializing M matching tuples from a relation with N tuples, and therefore $mr = M/N$ is called the match ratio. Suppose we partition based on 16 radix bits, which requires histogram construction and 2 invocations of \textsc{radix\_partition}. Similar to the model for the sort-based group-by, we estimate the bandwidth of random global memory accesses by $B_G/24$ as shown in Section~\ref{sec:random-access-cause}. Therefore, the cost of case 1 is 
$$C_1 = \underbrace{\frac{2\times 4M}{B_G}}_{\text{Read IDs and write output}} + \underbrace{\frac{4M}{B_G/24}}_{\text{Read input column}} = \frac{104M}{B_G}.$$
The cost of case 2 is
$$C_2 = \underbrace{\frac{4N + 2\times 4\times 4N}{B_G}}_{\text{Partition cost}} + \underbrace{\frac{3\times 4M}{B_G}}_{\text{Gather cost}}= \frac{36N+12M}{B_G}.$$

To make $C_1 > C_2$, we need $mr > 0.39$. In the actual implementation, however, the GFTR-based join is often better than the GFUR-based even for $mr < 0.39$. That is because while GFUR partitions the (keys, tuple IDs) pair in the transformation phase, GFTR partitions (keys, 1st payload column) pair making the partitioning of the first payload column free. 
For instance, when materializing two payload columns, GFTR is better than GFUR for $mr > 0.2$ based on the model.

\subsection{Optimized Hash-based Group-by}\label{sec:opt-hash-groupby}
We have explained how the hash-based group-by in cuDF works in Section~\ref{sec:hash-group-by}. In the experiments, we will show that the hash-based group-by unavoidably performs poorly for a large number of groups due to random hash table accesses. Nonetheless, we argue that there is still a great space for improvement for a small to medium number of groups, where the current cuDF implementation mainly suffers from contention due to concurrent updates on the aggregate values. The concurrent updates using atomic operations serialize the memory accesses and hurt parallelism. As the number of groups grows larger, less contention happens due to more spread-out accesses, and therefore the performance becomes better. 

To alleviate the contention, we propose to use the \emph{shared memory} for per-thread-block aggregation and then update the final aggregate results with the local results of each thread block. In doing so, we can reduce the number of global atomic operations and leverage much faster shared memory atomic operations. 

Simple as it seems, to apply this optimization requires a redesign of the hash-based group-by. There are two reasons. Firstly, cuDF aggregates values into sparse arrays, each having as many entries as the input. Such large arrays exceed the capacity of the shared memory. Therefore, we must store aggregate values in \emph{dense} arrays, which requires us to assign each group a unique ID from $\{0,...,g-1\}$. Secondly, \emph{before} aggregating, we must already know the number of groups in order to tell if the shared memory is sufficient in capacity. It is only possible for cuDF to know this after the aggregation since it aggregates while inserting into the hash table. 

To apply this optimization, we redesign the hash-based group-by as shown in Algorithm~\ref{alg:opt-hash-group-by}. The core of the redesign is to associate each input tuple with a unique ID from $\{0,...,g-1\}$ \emph{before} the aggregation so that we can aggregate into \emph{dense} arrays, which can then be buffered by the shared memory. The new algorithm works in two stages. In the first stage, we insert each key $k$ and its tuple ID $i$ into the global hash table. If the insertion succeeds, then we increment the number of groups by 1 and write the current number of groups into \texttt{groupId[i]} as the group ID. No matter if the insertion succeeds or not, we store the tuple IDs with which $k$ is inserted into \texttt{groupIdLoc[i]}. In the second stage, we first check if we should apply the optimization by checking whether the shared memory is sufficient given the number of groups. If yes, we first retrieve the the input tuple with ID $i$ its group membership by $\texttt{groupId[groupIdLoc[i]]}$ and then aggregate into the corresponding location in the shared memory. After the thread block has finished the aggregation, it then aggregates its local results into the global results. If the shared memory is not sufficient, we fall back to directly aggregating into the global memory. 

\SetKwInput{KwInput}{Input}
\SetKwInput{KwOutput}{Output}
\newcommand\mycommfont[1]{\ttfamily\textcolor{blue}{#1}}
\SetCommentSty{mycommfont}
\SetKwComment{Comment}{$\triangleright$\ }{}

\begin{algorithm}[t]
\small
\caption{Optimized Hash-based Group-by}\label{alg:opt-hash-group-by}
\KwInput{Relation $R(k,r_1,\dots,r_n)$, Agg functions $F=[f_1,\dots f_n]$}
\KwOutput{Relation $A(k,a_1,\dots,a_n)$}
\BlankLine

\SetKwFunction{FMain}{Main}
\SetKwProg{Fn}{Function}{:}{}
\Fn{\FMain{$R$, $F$, $A$}}{
    \Comment{Stage 1 - Assign Group IDs}
    $HT \gets \text{HashTable}(|R|)$\;
    $groupId \gets \text{alloc\_on\_device}(int, $|R|$)$\;
    $groupIdLoc \gets \text{alloc\_on\_device}(int, $|R|$)$\;
    $numGroups \gets 0$\;
    \texttt{compute\_group\_id}$(R.k, HT, groupId, groupIdLoc, numGroups)$\;%
    \Comment{Stage 2 - Aggregate}
    \texttt{compute\_agg}$(R, F, groupId, groupIdLoc, A, numGroups)$\;%
}

\SetKwFunction{FCGI}{compute\_group\_id}
\Fn{\FCGI{$key$, $HT$, $gId$, $gIdLoc$, $N$}}{
    \For{each thread $i$}{%
        \Comment{Insert or return the value if key has existed}
        $status, gIdLoc[i] \gets HT.\text{insert\_and\_find}(key[i],i)$\;
        \uIf{$status \text{ is Success}$} {
            $gId[i] \gets \text{atomicAdd}(N, 1)$\;
            $A.k[gId[i]] \gets R.k[i]$\;
        }
    }
}

\SetKwFunction{FCA}{compute\_agg}
\Fn{\FCA{$R$, $F$, $gId$, $gIdLoc$, $A$, $N$}}{
    \uIf{$\text{is\_smem\_enough}(R, F, N)$} {%
        $\_\_shared\_\_$ $G(g_1,\dots,g_n)$\;
        \For{each thread $i$}{%
            $g \gets gId[gIdLoc[i]]$\;
            \For{each $f_j\in F$}{%
                $f_j.\text{update}(G.g_j[g], R.r_j[i])$\;
            }
        }
        synchronize the thread block\;
        \For{each group $g$} {
            \For{each $f_j\in F$}{%
                $f_j.\text{update}(A.a_j[g], G.g_j[g])$\;
            }
        }
    }
    \Else {%
       Aggregate directly into $A$. Omitted\;
    }
}
\end{algorithm}

\subsection{Partition-based Group-by}\label{sec:pgb}
Inspired by the superior performance of the partition-based join~\cite{sioulas19-partitioned-radix-join}, we propose a partition-based group-by and demonstrate that partitioning the data can also substantially benefit the group-by. 
The partition-based group-by can be implemented with GFUR or GFTR. 
For brevity, we explain how the GFTR version works. 

Similar to the sort-based group-by, this algorithm also works in a transform-then-group way. 
In the transformation step, we partition the group keys and \emph{all} payload data using the same partitioning procedure as our partitioned hash join (see Section~\ref{sec:opt-phj}). 
In the grouping step, we assign each partition to a thread block. 
Within the thread block, grouping and aggregating resemble cuDF's hash-based group-by except that the thread block maintains a hash table in its \emph{shared memory} to achieve faster random accesses. 
Similar to cuDF, the hash table does not contain the actual aggregated values but only the necessary information to locate where those values are stored. 
On the other hand, our implementation differs from cuDF by decoupling the aggregation with the assignment of group membership. 
Instead of inserting into the hash table (membership assignment) and \emph{immediately} aggregating the payload data, we insert all keys before performing aggregation in order to know the number of groups and the offset of each group in the output relation. 
This approach has two major benefits. 
Firstly, knowing the offset into the output relation allows us to store the aggregated values in a \emph{dense} array instead of a sparse array to save memory consumption and the cost of gathering. 
Secondly, knowing the number of groups within the partition enables choosing the most optimal aggregation procedure. There are three cases to be considered. 
(1) When there is only one group within the partition, we can aggregate the data using the highly optimized block-wide reduction procedure~\cite{cub}. 
(2) When the number of groups is relatively small, we can allocate a buffer in the shared memory and assign a slot for each group and each aggregation function. Then, we first aggregate the data into the shared memory and then flush the aggregated values to the global memory. 
(3) When the number of groups is so high that the shared memory is not sufficient, we can directly aggregate the data into the global memory. 
Optimizing for the first two cases are especially important for the partition-based group-by since the number of groups in a partition is expected to be much smaller than that of the entire data. 

Partition-based group-by faces two major challenges. 
The first one is that partitions may vary in sizes causing uneven workload distribution among thread blocks. 
If this skew is caused by the poor choice of radix bits, we can alleviate it by partitioning based on the \emph{hashed} group keys. 
Hashing the group keys will not cause significant overhead since it is required in the grouping step anyway. 
On the other hand, if the skew is caused by the affluence of identical group keys (e.g., low number of groups), the solution is to apply sort-based or hash-based group-by for the overly large partitions. 
The second challenge is that the hash table may run out of the shared memory if a partition contains many groups. The solution is to limit the sizes of partitions. Suppose the shared memory can accommodate a hash table for $M$ groups at maximum, then we make sure to produce enough partitions so that each partition does not have more than $M$ tuples. 

\subsection{Faster Sorting via Dictionary Encoding}\label{faster-sorting}
In Section~\ref{sort-groupby-gftr}, we show how the GFTR technique uses more sorting to reduce random accesses in the sort-based group-by, and here we present how to use the dictionary encoding to further reduce the sorting costs. When there are many different payload columns to aggregate, sorting all pairs of key and payload columns becomes very costly. The cost of sorting is even higher for keys that are more than 4-byte long since the complexity of radix sorting grows with the key length. Although partition-based or hash-based group-by may have better performance in these cases, sort-based group-by supports more aggregation functions and more importantly can often save the following order-by in many SQL queries (around 7 out of 22 TPC-H queries). 

To reduce the cost of sorting, we can encode the group keys with numbers from domain $\{0,...,g-1\}$ ($g$ is the number of groups). The encoded keys have at most ($\log_2 g$) non-zero bits, and sorting the encoded keys only takes $\lceil(\log_2 g)/8\rceil$ passes. We already know the number of groups from aggregating the first payload column using normal sorting. For the remaining payload columns, we can sort and aggregate them with \emph{encoded} keys. 

To understand when this optimization pays off, we develop the following mathematical model. Assuming that the number of groups $g$ is known, we compare two scenarios -- (1) sorting $p$ different 4-byte payload columns with $k$-byte keys in $k$ passes ($k \geq 4$) and (2) encoding the $k$-byte keys into 4-byte keys and then sorting $p$ 4-byte payload columns using only ($\log_2 g$) bits of the encoded keys in $\lceil(\log_2 g)/8\rceil$ passes. Let $N$ be the number of input tuples, $D(N,g)$ be the cost of encoding $N$ keys using a dictionary with $g$ alphabets. Then, the cost of (1) is

$$C_1 = p\cdot \frac{kN + k\cdot N\cdot 2(k+4)}{B_G}.$$

And the cost of (2) is

$$C_2 = D(N,g)+p\cdot \frac{4N + \lceil(\log_2 g)/8\rceil \cdot N\cdot 2(4+4)}{B_G}.$$

We assume $k=4$ from now on. 
To make $C_2 < C_1$, we need 
$$D(N,g) < \frac{16p(4-\lceil(\log_2 g)/8\rceil)}{B_G}N.$$
Next, we analyze the cost of dictionary encoding, namely $D(N,g)$. Suppose the non-empty entries of the hash table can fit into a certain type of memory (e.g., global memory, L2 cache, L1 cache) whose bandwidth is $B_M$. $D(N,g)$ consists of the building cost and probing cost and can be calculated as
\begin{equation*}
\begin{split}
D(N,g) & = \underbrace{\bigg(\frac{kg}{B_G}+\frac{(k+4)g}{B_M/\min(g,24)}\bigg)}_{\text{Build cost}} + \underbrace{\bigg(\frac{kN+4N}{B_G}+\frac{(k+4)N}{B_M/\min(g,24)}\bigg)}_{\text{Probe cost}} \\
& \overset{k=4}{=}  \frac{4g+8N}{B_G}+\frac{8(N+g)}{B_M/\min(g,24)}.
\end{split}
\end{equation*}
Remember we use $B_M/24$ to approximate the random access throughput. For A100 GPUs, the L1 and L2 bandwidth are around $12B_G$ and $4.5B_G$, respectively~\cite{dissect-a100}. Assuming $N=2^{28}$, we provide Table~\ref{tab:de-opt} to illustrate for different numbers of groups, what number of payload columns $p$ can make $C_2<C_1$ for an A100 GPU. From the table, we see that when $k=4$ and $g < 2^{15}$, two payload columns in total will already make the optimization pay off. 

\begin{table}[]
\centering
\caption{Conditions for benefitting from dictionary encoding.}
\label{tab:de-opt}
\begin{subtable}[h]{0.3\columnwidth}
\centering
\caption{$k=4$}%
\label{}
\begin{tabular}{ll}
\hline
$g$                & For $C_2 < C_1$ \\ \hline
$[2^0, 2^{14}]$    & $p \ge 1$       \\
$[2^{15}, 2^{16}]$ & $p \ge 3$       \\
$[2^{17}, 2^{22}]$ & $p \ge 5$       \\
$[2^{23}, 2^{24}]$ & $p \ge 14$      \\ \hline
\end{tabular}%
\end{subtable}
\begin{subtable}[h]{0.3\columnwidth}
    \centering
    \caption{$k=8$}%
    \label{}
    \begin{tabular}{ll}
    \hline
    $g$                & For $C_2 < C_1$ \\ \hline
    $[2^0, 2^{21}]$    & $p \ge 1$       \\
    $[2^{22}, 2^{26}]$ & $p \ge 3$       \\
    $2^{27}$ & $p \ge 4$       \\
    $2^{28}$ & $p \ge 5$       \\ \hline
    \end{tabular}%
\end{subtable}
\end{table}

\section{Experimental evaluation}\label{sec:exp}
We evaluate the existing and optimized implementations on an A100 GPU. The GPU is based on the Ampere architecture and has 108 SMs, 192KB L1 cache, 40MB L2 cache, and 40GB global memory. The theoretical maximum memory bandwidth is 1555 GB/s. We run with CUDA version 12.2. 

\subsection{Microbenchmarks for Joins}
\subsubsection{Workload Description.}
We study the in-memory inner equi-join of two relations $R$ and $S$ on the GPU.
We assume that $R$ holds the primary keys, and $S$ holds the foreign keys.
By default, we assume the keys and payloads in both relations are all 4-byte integers.
In one of our experiments, we evaluate the performance when keys and payloads
are 8-byte long. With the notation $1\text{G}\bowtie 2\text{G}$, we mean that we
join two relations whose total sizes are 1 GB and 2 GB respectively, including payloads.
We generate keys that take values from $0$ to $|R|-1$, and we randomly shuffle
them.

\subsubsection{Measurements.}
Following previous
work~\cite{sioulas19-partitioned-radix-join,Lutz22-tritonjoin,Balkesen13-partitioned-hash-join}, the throughput of a join is defined by $\frac{|R|+|S|}{\text{total
time}}$. We also report the time spent in the three join phases (transformation, match finding, and materialization). For GFTR-optimized implementations, we count the sorting/partitioning of the payload data as materialization time instead of transformation time. All experiments are repeated at least seven times, and we report the median.

\subsubsection{Implementations.}
We evaluate the following algorithms:
\begin{itemize}[wide, labelwidth=!, labelindent=0pt]
    \item[\textbf{SMJ-UR} (SU)] Sort-merge join with GFUR (Section~\ref{sec:unopt-smj}).
    \item[\textbf{PHJ-UR} (PU)] Partitioned hash join with GFUR (Section~\ref{sec:unopt-phj}).
    \item[\textbf{cuDF}] Non-partitioned hash join from cuDF (Section~\ref{sec:nphj}).  
    \item[\textbf{SMJ-TR} (ST)] Sort-merge join with  GFTR (\textbf{Ours}).
    \item[\textbf{PHJ-TR} (PT)] Partitioned hash join GFTR (Section~\ref{sec:opt-phj}) (\textbf{Ours}).
\end{itemize}
For brevity, we use wildcards for a class of implementations. For example, ``SMJ-*'' refers to sort-merge join implementations with GFUR or GFTR, and ``*-TR'' covers any implementations with GFTR.

\subsubsection{Clustered vs. Unclustered \textsc{Gather}}
In the first experiment, we check how transforming all columns by sorting or partitioning
before any \textsc{gather}s alters performance.
As explained in Section~\ref{sec:opt}, the \textsc{gather} primitive is used
in the materialization phase to write payload columns of matching tuples into the output relation.
Gathering from transformed columns increases the probability that threads in a
warp read payload values from the same cache line.

\begin{figure}[t]
    \centering        
    \includegraphics[width=0.5\linewidth]{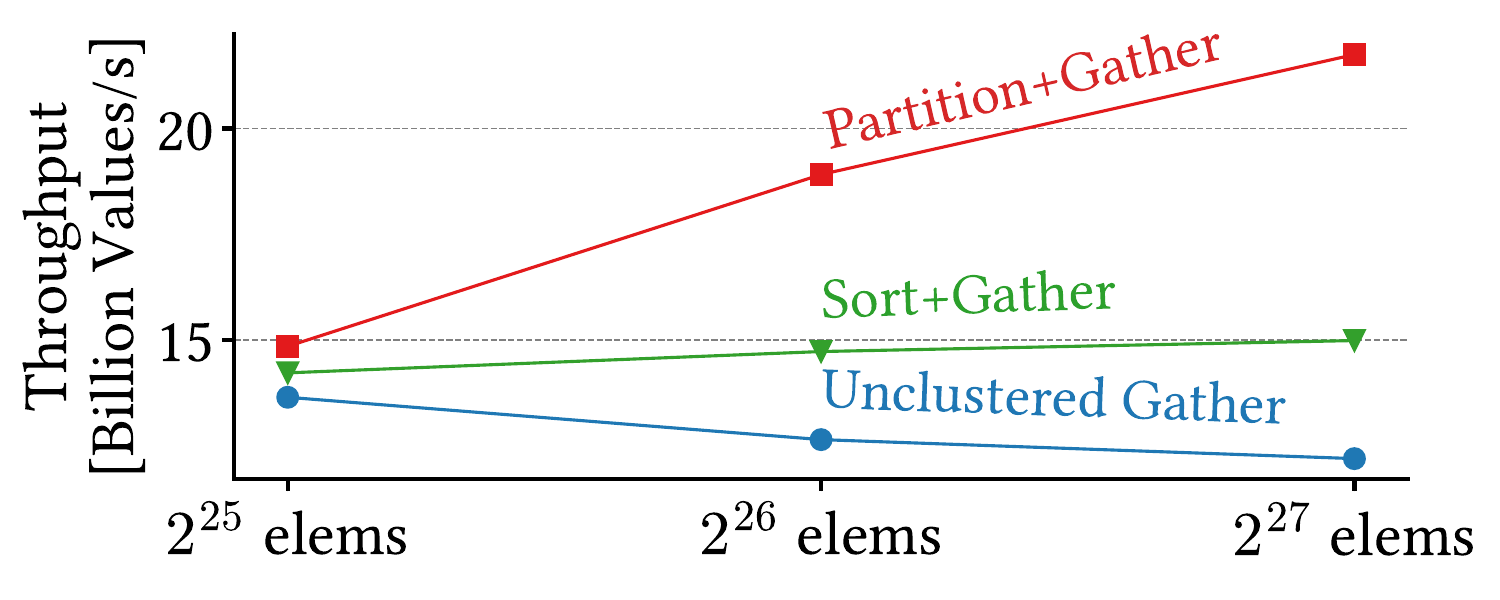}
    \caption{Unclustered vs. clustered \textsc{gather} with extra costs.}
    \label{fig:seq-vs-random}
\end{figure}

We examine the differences in performance between these two
\textsc{gather} operations with the added cost of sorting or partitioning.
Figure~\ref{fig:seq-vs-random} shows the throughput of \textsc{gather}s used
in *-UR, SMJ-TR, and PHJ-TR.
Even with the sorting and partitioning overhead, the clustered
\textsc{gather} outperforms the unclustered \textsc{gather}.
Partitioning and a clustered \textsc{gather} are
1.79x faster than an unclustered \textsc{gather}. 
For sorting and a clustered \textsc{gather}, the speedup is 1.23x. 
Therefore, the GFTR pattern
reduces the materialization cost despite the additional
transformation cost.

\subsubsection{Narrow and Wide Join Performance}\label{sec:mb_q2}
We now compare the performance of narrow and wide joins. We first assume a match ratio of 100\%, meaning
all tuples in $S$ have a match partner in $R$. Moreover, $S$ has twice as many
tuples as $R$, namely $|S|=2|R|$.

\begin{figure*}[t]
    \begin{minipage}[t]{.48\linewidth}
        \centering
        \includegraphics[width=1\linewidth]{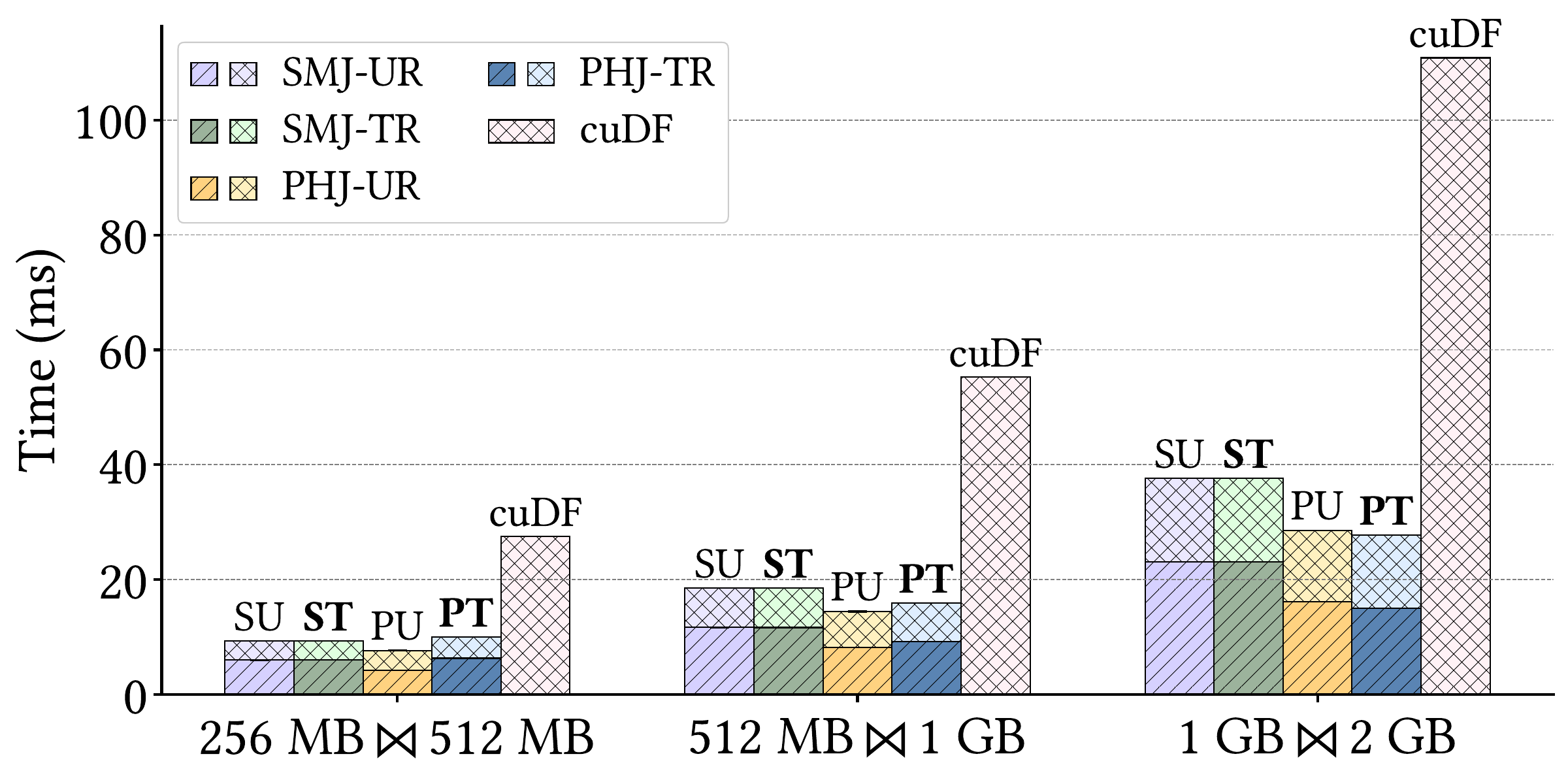}%
        \caption{Time breakdown of narrow joins.}
        \label{fig:narrow-table-breakdown}
    \end{minipage}\hfill
    \begin{minipage}[t]{.48\linewidth}
        \centering
        \includegraphics[width=1\linewidth]{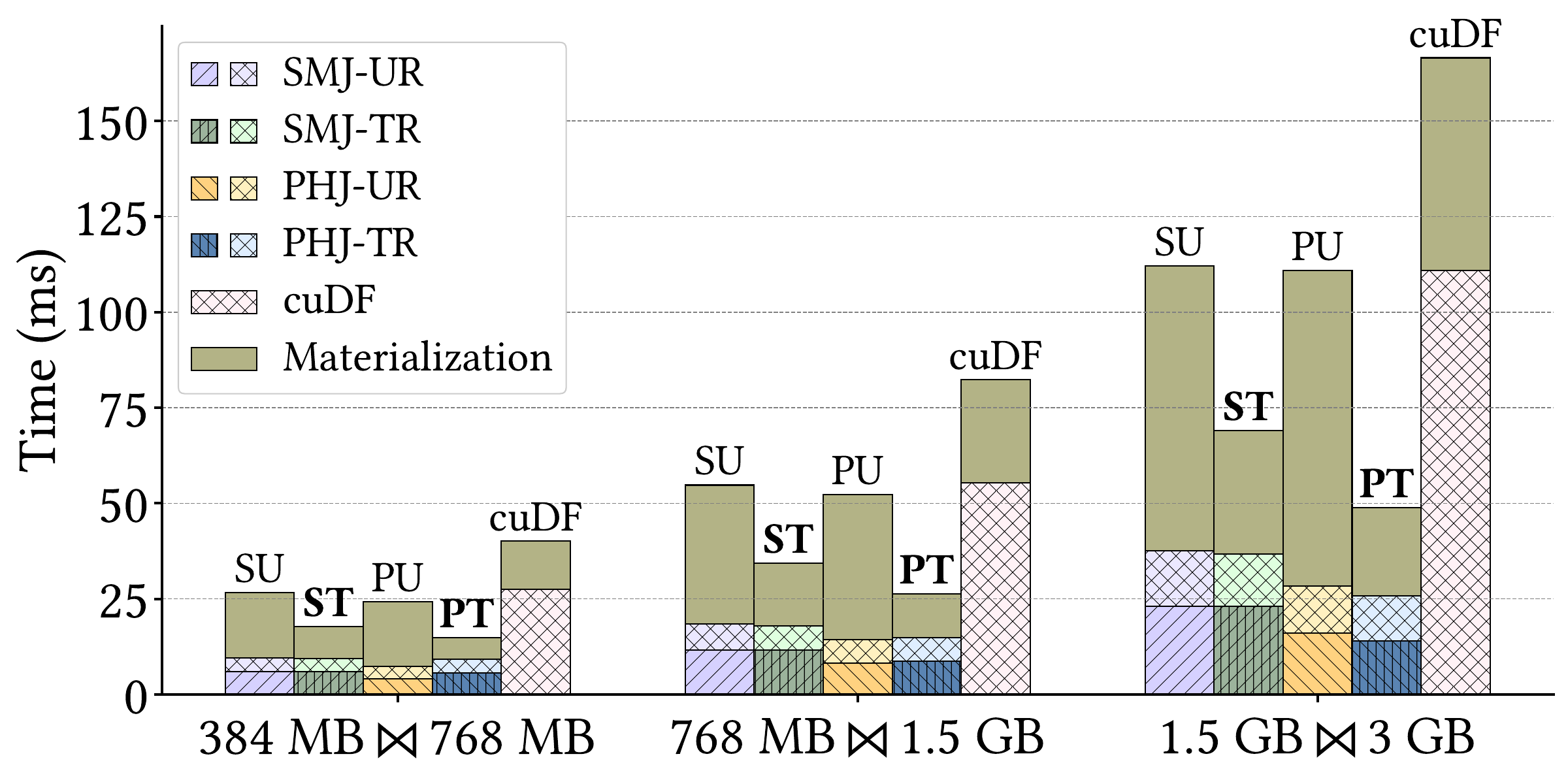}%
        \caption{Time breakdown of wide joins.}
        \label{fig:wide-table-breakdown}
    \end{minipage}\hfill
\end{figure*}

(\textbf{Narrow joins}) We depict the time breakdown of different GPU-based
implementations for processing narrow joins (Figure~\ref{fig:narrow-table-breakdown}). A bar in the figure corresponds to an implementation and consists of the transformation time (bottom) and match-finding time (top). cuDF does not do transformation. 
The GPU-accelerated sort-merge join and partitioned hash join are up to 4 times faster than the cuDF join. 
We also observe that PHJ-* performs better than SMJ-* for joining
narrow tables. Since the joins are narrow,
SMJ-TR is identical to SMJ-UR and performs worse than PHJ-*.
PHJ-UR performs slightly better than PHJ-TR for smaller input sizes, yet both have almost identical performance for $1\text{G}\bowtie 2\text{G}$.
Overall, the performance of PHJ-UR and PHJ-TR is very close. This indicates that the radix-sort-based partitioning performs as well as the bucket-chaining-based one. cuDF is the least inefficient for this workload due
to the random accesses during the building and probing.

(\textbf{Wide joins}) To evaluate wide joins, we append to each relation two payload columns, (Figure~\ref{fig:wide-table-breakdown}). Even with two payload
columns per relation, the materialization time dominates most *-UR implementations.
Thus, *-TR has a significant advantage
over *-UR. More specifically, SMJ-TR is
approximately 1.6x faster than SMJ-UR, and although sorting is considered
costly, SMJ-TR achieves a 1.6x speedup over PHJ-UR. PHJ-TR stands out as the
most performant algorithm, achieving around 2.3x and 1.4x better
performance than PHJ-UR and SMJ-TR, respectively. It is better than SMJ-TR
mainly because the \textsc{radix-partition} needs fewer passes for partitioning.
The non-partitioned hash join (cuDF) still performs the worst for the wide join case; however, it has a lower materialization cost than *-UR since materializing the probe table is clustered.

\begin{figure*}[t]
    \begin{minipage}[t]{.48\linewidth}
        \centering
        \includegraphics[width=\linewidth]{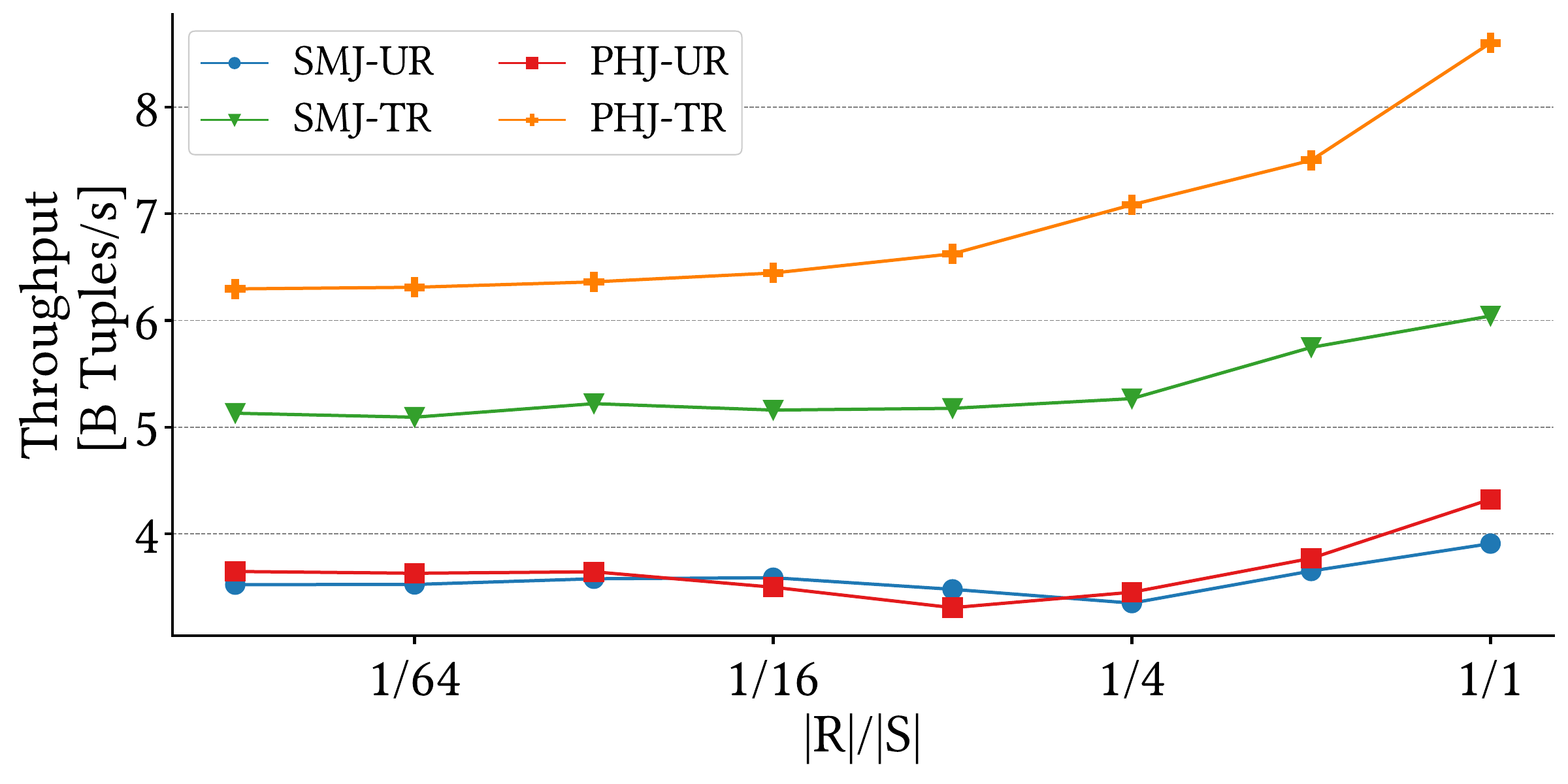}%
        \caption{Effect of $R|/|S|$.}
        \label{fig:wide-table-ratio}
    \end{minipage}\hfill
    \begin{minipage}[t]{.48\linewidth}
        \centering
        \includegraphics[width=\linewidth]{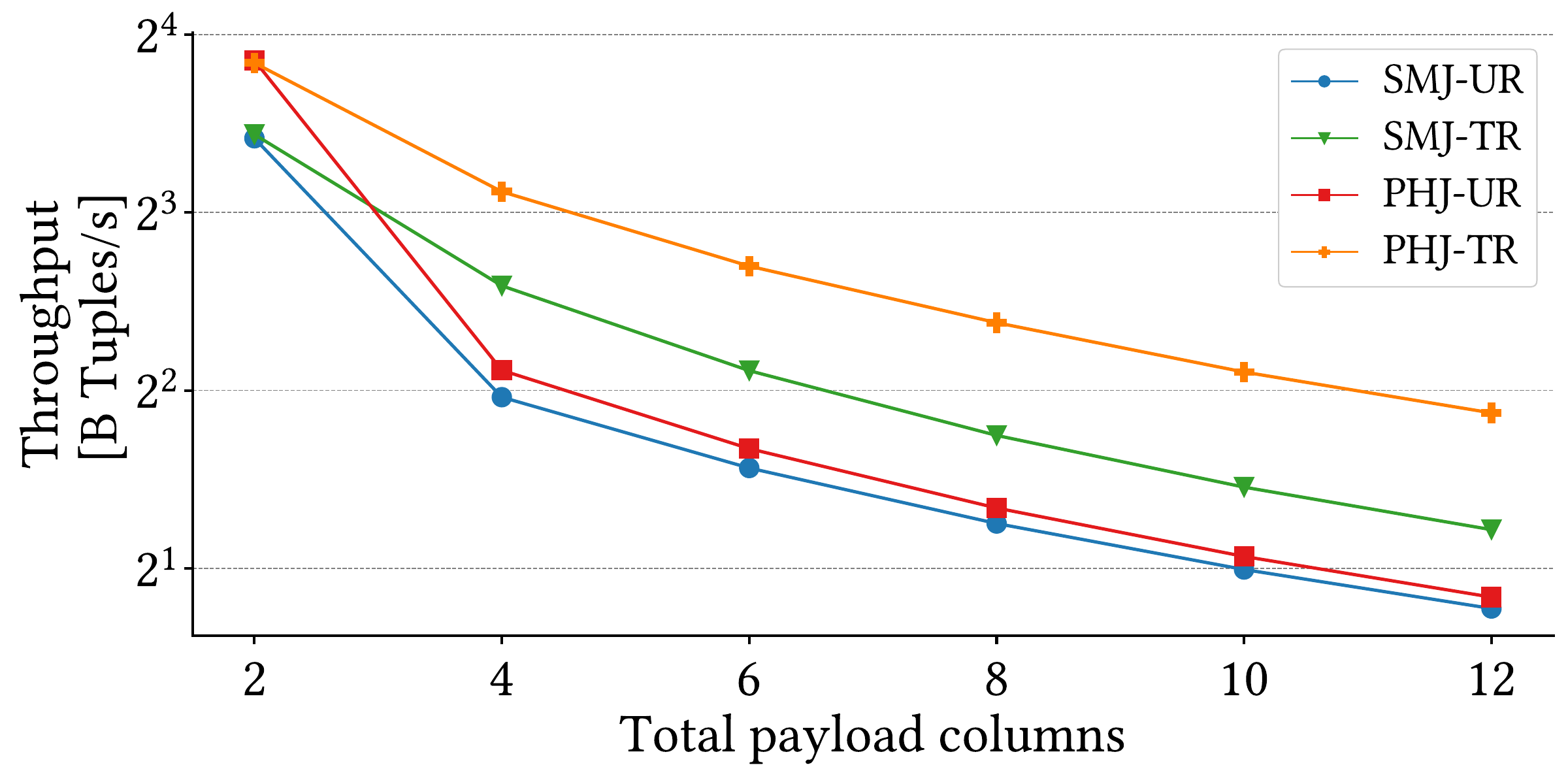}%
        \caption{Effect of number of payloads.}
        \label{fig:wide-table-cols}
    \end{minipage}\hfill
\end{figure*}

We evaluate two more factors that can affect the efficiency of wide joins. Figure~\ref{fig:wide-table-ratio} studies the relative size between $R$ and $S$ ($|S|=2^{27}$). We observe that *-TR still outperforms *-UR even though the materialization cost is lower when $R$ is small. Additionally, Figure~\ref{fig:wide-table-cols} studies the effect of number of payload columns ($|R|=|S|=2^{27}$). As the number of payload columns increases, PHJ-TR and SMJ-TR maintain 2x and 1.3x speedups over PHJ-UR, respectively. 


\subsubsection{Match Ratio}
We now study the effect of different match ratios.
With a high match ratio, more data are materialized, and
therefore *-TR implementations perform better. On the other hand, a low match
ratio, makes the materialization using unclustered
\textsc{gather}s less costly because only a small fraction of payload data are
materialized. We adjust the match ratio by replacing a corresponding fraction
of primary keys with non-matching values.

\begin{figure*}[t]
    \begin{minipage}[t]{.48\linewidth}
        \centering
        \includegraphics[width=\linewidth]{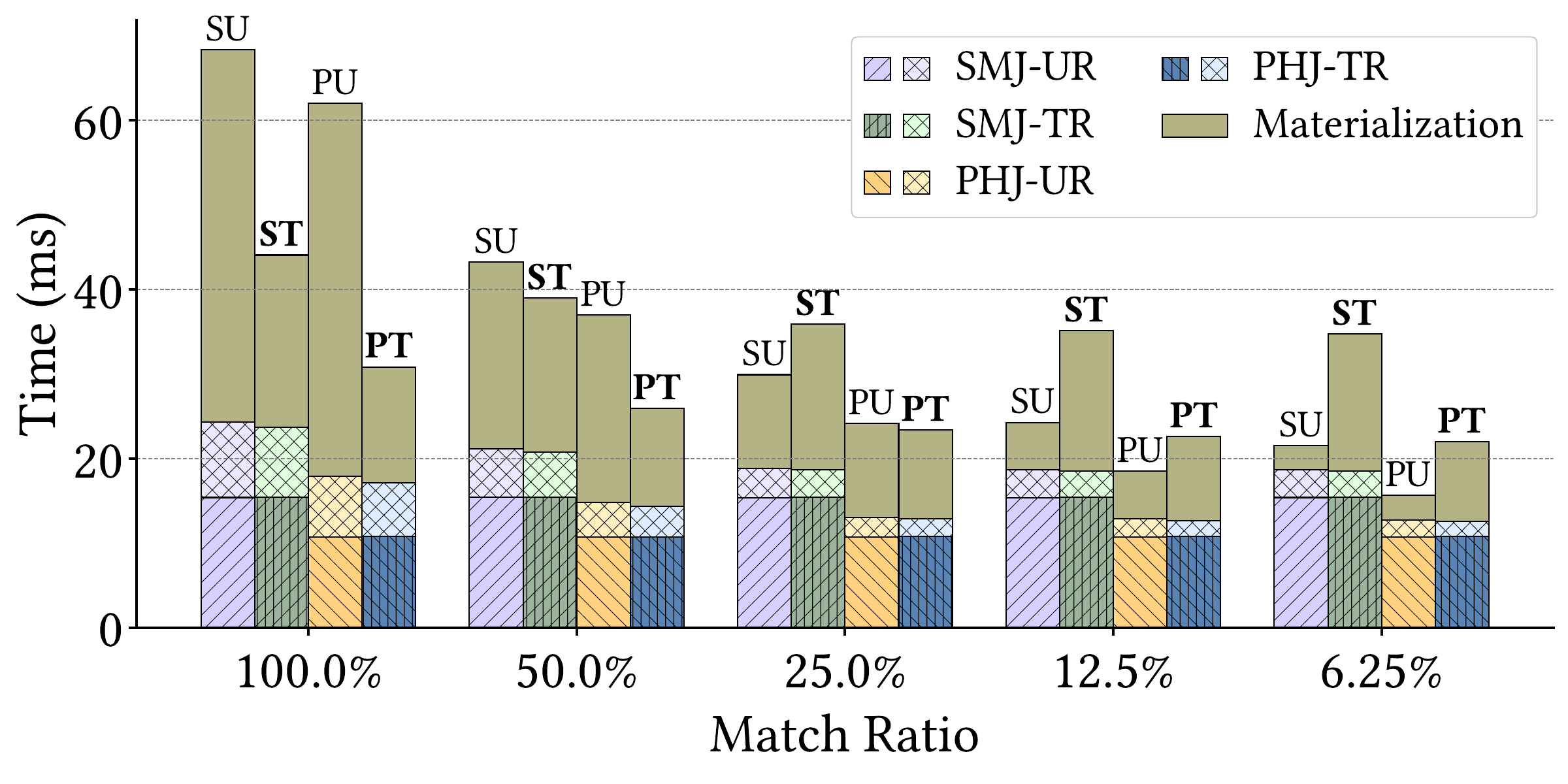}%
        \caption{Different match ratios.}%
        \label{fig:match-ratio}
    \end{minipage}\hfill
    \begin{minipage}[t]{.48\linewidth}
        \includegraphics[width=\linewidth]{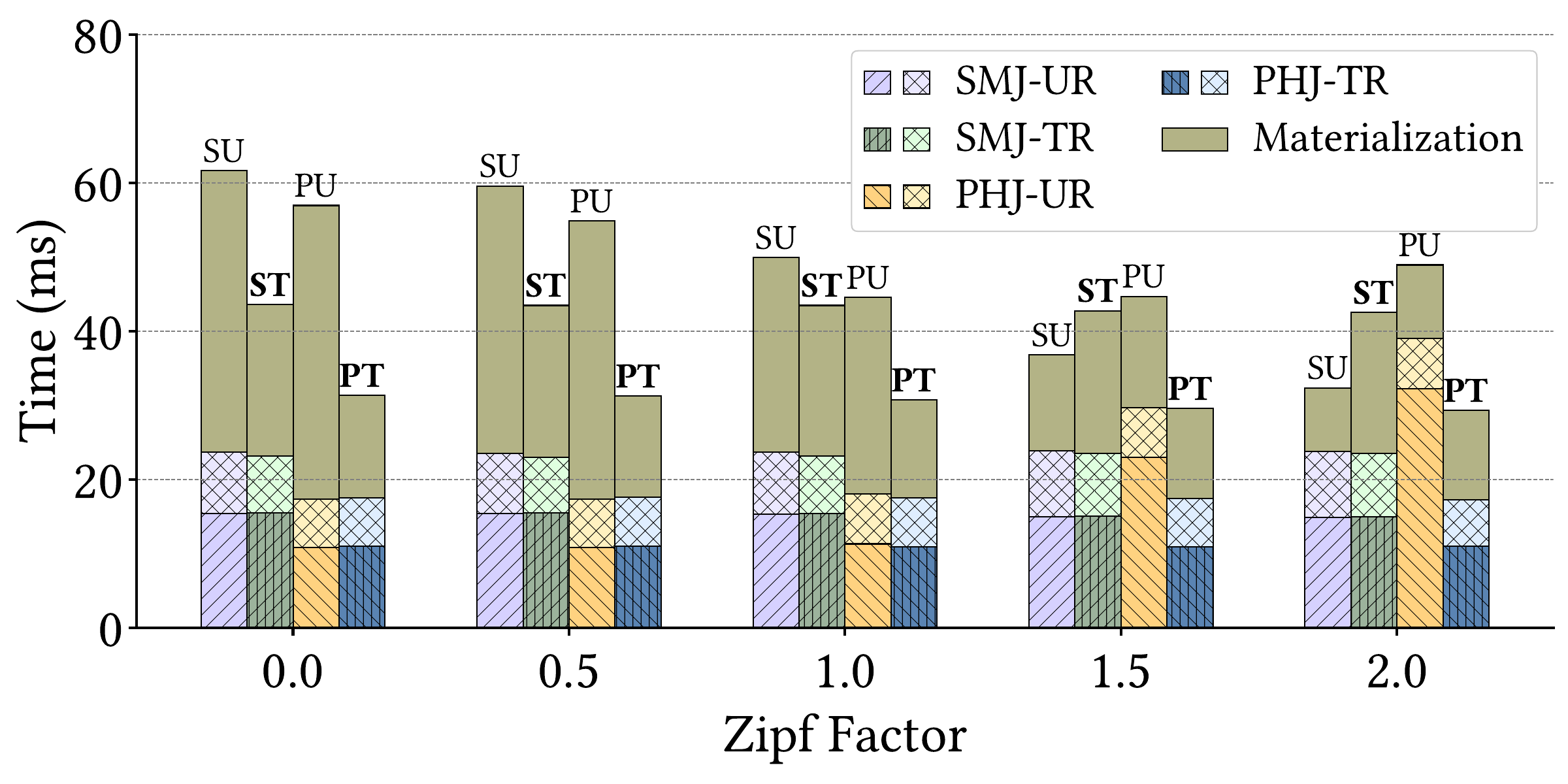}%
    \caption{Foreign key skewness.}%
    \label{fig:data-skewness}
    \end{minipage}\hfill
\end{figure*}

We show the effect of the match ratio
in processing $1.5\text{G}\bowtie 1.5\text{G}$ in Figure~\ref{fig:match-ratio}
for relations $R, S$ with two payload columns each ($|R|=|S|=2^{27}$).
We observe that *-TR implementations have better performance for high match
ratios, and when the ratio drops below $25\%$, *-TR is slower
because the materialization is no longer a bottleneck. The empirical threshold 25\% is close to our model analysis in Section~\ref{sec:opt-phj}.
PHJ-UR performs the best when the match ratio
is low, due to the low materialization needs. 

{\color{ETHc} Figure~\ref{fig:phj-mr} shows the measured and model-predicted (see Section~\ref{sec:opt-phj}) time differences between PHJ-UR and PHJ-TR. A positive value means PHJ-TR is faster. The figure shows that our model not only successfully predicts 25\% to be the turning point but accurately captures their relative differences for other match ratios.}

\begin{figure}[H]
    \centering
    \resizebox{0.55\linewidth}{!}{\begin{tikzpicture}
        \begin{axis}[
            title={},
            xlabel={Match ratio},
            ylabel={Time difference (ms)},
            xmin=0, xmax=6,
            ymin=-10, ymax=35,
            xtick={1,2,3,4,5},
            xticklabels={$100\%$,$50\%$,$25\%$,$12.5\%$,,$6.25\%$},
            ytick={-10,0,10,20,30},
            legend pos=north east,
            ymajorgrids=true,
            axis x line=bottom,        
            axis y line=left,          
            height=4cm,
            width=10cm
        ]
        
        \definecolor{mygreen}{rgb}{0.0, 0.6, 0.3}
        
        \addplot[
            color=blue,
            dashed,
            mark=*,
            ]
            coordinates {
(1,25.5488408282958)(2,9.66712896205788)(3,1.72627302893891)(4,-2.24415493762058)(5,-4.22936892090032)
            };
        
        \addlegendentry{\texttt{Estimated}}
        
        \addplot[
            color=blue,
            mark=*,
            ]
             coordinates {
(1,30.4055)(2,10.5541)(3,0.5844)(4,-4.36836)(5,-6.52067)
            };
        \addlegendentry{\texttt{Real}}
        \end{axis}
\end{tikzpicture}}%
    \caption{\color{ETHc} Time(PHJ-UR) - Time(PHJ-TR)}
    \label{fig:phj-mr}
\end{figure}

\subsubsection{Data Skew}
Many datasets have a skewed foreign-key distribution which causes unbalanced
workload distribution among execution units (e.g., thread blocks)
during join processing. The unbalanced distribution reduces the level of
parallelism by assigning a disproportionately large amount of the work to an
execution unit. Following previous work~\cite{Balkesen15-hashjoin}, we
generate the foreign key based on the Zipfian distribution and vary the Zipf
factor to adjust the level of skewness. A high Zipf factor means that the data
distribution is centralized at only a few values. Once again, $R$ and $S$ contain
two payload columns each, and they both have a size of $1.5$ GB ($|R|=|S|=2^{27}$). 
As shown in Figure~\ref{fig:data-skewness}, PHJ-UR is particularly sensitive to data skewness, with the partitioning cost
increasing significantly as the Zipf factor grows and exceeds $1$. This is mainly because of the uneven workload distribution among threads when allocating and bookkeeping the buckets in the first and second pass of partitioning, which leads to a high synchronization cost within a thread block. 
This performance degradation indicates that the current bucket
chaining implementation is unsuitable for non-uniform data, a bottleneck also
found by Sioulas et al.~\cite{sioulas19-partitioned-radix-join}. 

On the other hand, the \textsc{radix-partition} primitive used in PHJ-TR
and SMJ-* is more robust to data skewness and exhibits consistent performance
across all Zipf factors. \textsc{radix-partition} ensures all threads get the same amount 
of work regardless of the data distribution. In all implementations, the match-finding phase is
robust against data skewness, albeit for different reasons. In SMJ-*, the Merge
Path algorithm helps balance the workload of each GPU thread. In PHJ-*,
as discussed in Section~\ref{sec:unopt-phj} and~\ref{sec:opt-phj}, we perform a
load-balancing step to decompose large partitions before the match-finding phase.
As the foreign keys become increasingly skewed, the materialization cost is
reduced because only a small portion of primary keys have matches. When the Zipf
factor exceeds $1$, SMJ-UR becomes more competitive due to its consistent sorting
performance and lower materialization costs. In all cases, PHJ-TR has the best performance of all.

\subsubsection{Data Types}\label{sec:data_types}
Joins in the real world involve keys and payloads of various data types,
whose sizes range from one byte to an arbitrary number of bytes
(e.g., variable-length strings). Following previous work~\cite{Balkesen13-partitioned-hash-join,Bandle21-nonpartition-vs-partition}, we evaluate
the implementations with a mixture of 4-byte and 8-byte data. We let
$|R|=|S|=2^{27}$ tuples and we show the results in
Figure~\ref{fig:data-types}. With the introduction of 8-byte values into
payload columns (while keeping keys 4-byte), *-UR maintains nearly the same
performance as before, whereas *-TR experiences increased execution time due to
more costly sorting and partitioning. Because of the high sorting cost, SMJ-TR becomes less superior to *-UR.
In *-TR, the \textsc{radix-partition}
primitive reads and writes 4-byte keys and \emph{8-byte} payload values,
whereas in *-UR, it handles 4-byte keys and \emph{4-byte} tuple identifiers
regardless of the 8-byte payload values. When both keys and payload values are
8-bytes, the transformation and match-finding phase in all implementations 
become more expensive. SMJ-TR, in this case, has almost no advantage over *-UR. On the other hand, PHJ-TR leads the performance for all cases.
Notably, when the size of keys and payloads increases, the materialization
cost of *-UR methods remains nearly constant. This is because the unclustered
\textsc{gather} is bound by memory access latency, and accessing 4-byte and
8-byte values cause accesses to a similar number of cache lines.
For all combinations of data types examined here, PHJ-TR achieves the best
performance.

\subsubsection{Many-to-many joins}
\begin{figure*}[t]
    \begin{minipage}[t]{.45\linewidth}
        \includegraphics[width=\linewidth]{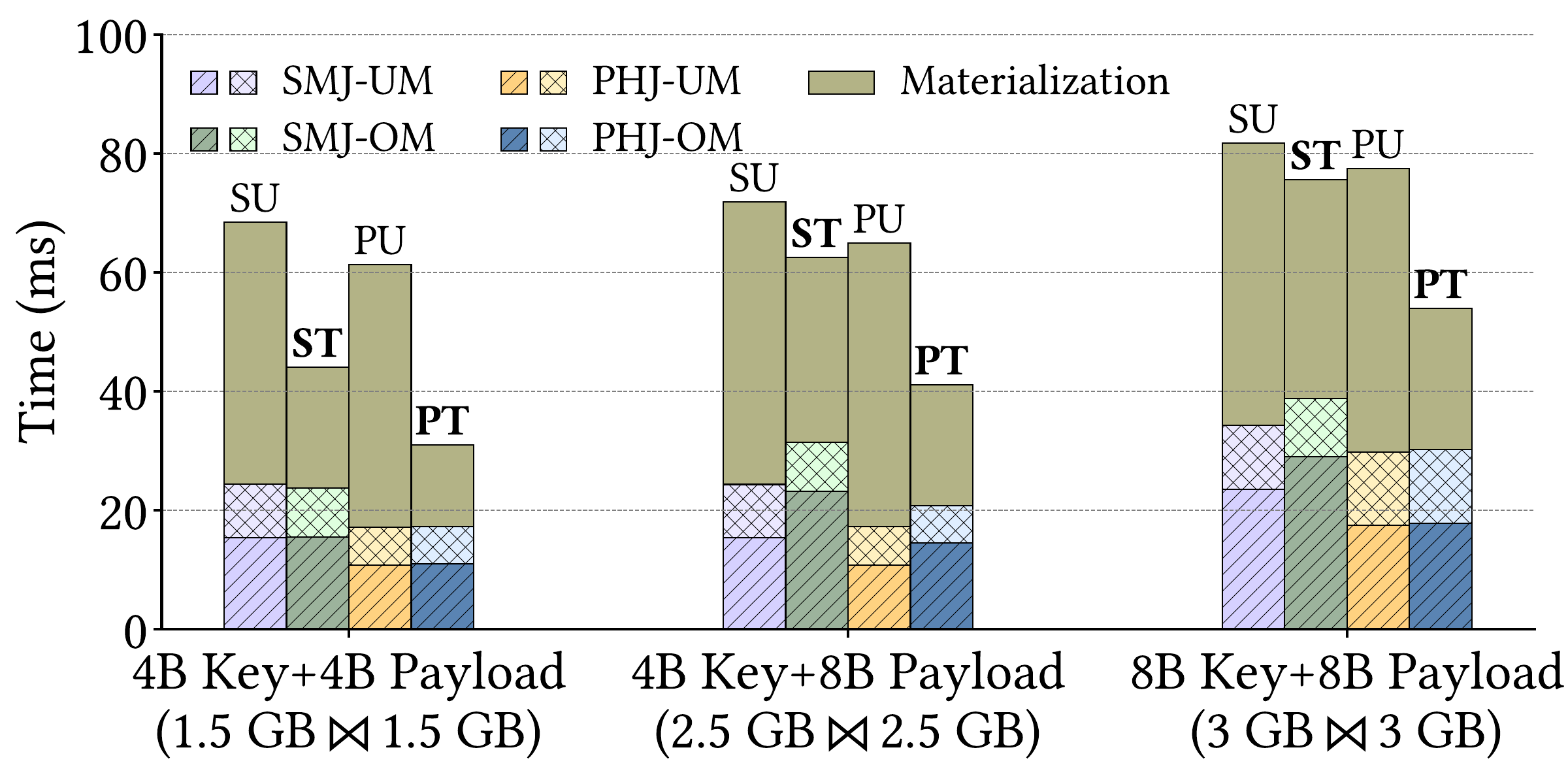}%
        \caption{Data types.}%
        \label{fig:data-types}
    \end{minipage}\hfill
    \begin{minipage}[t]{.45\linewidth}
        \includegraphics[width=\linewidth]{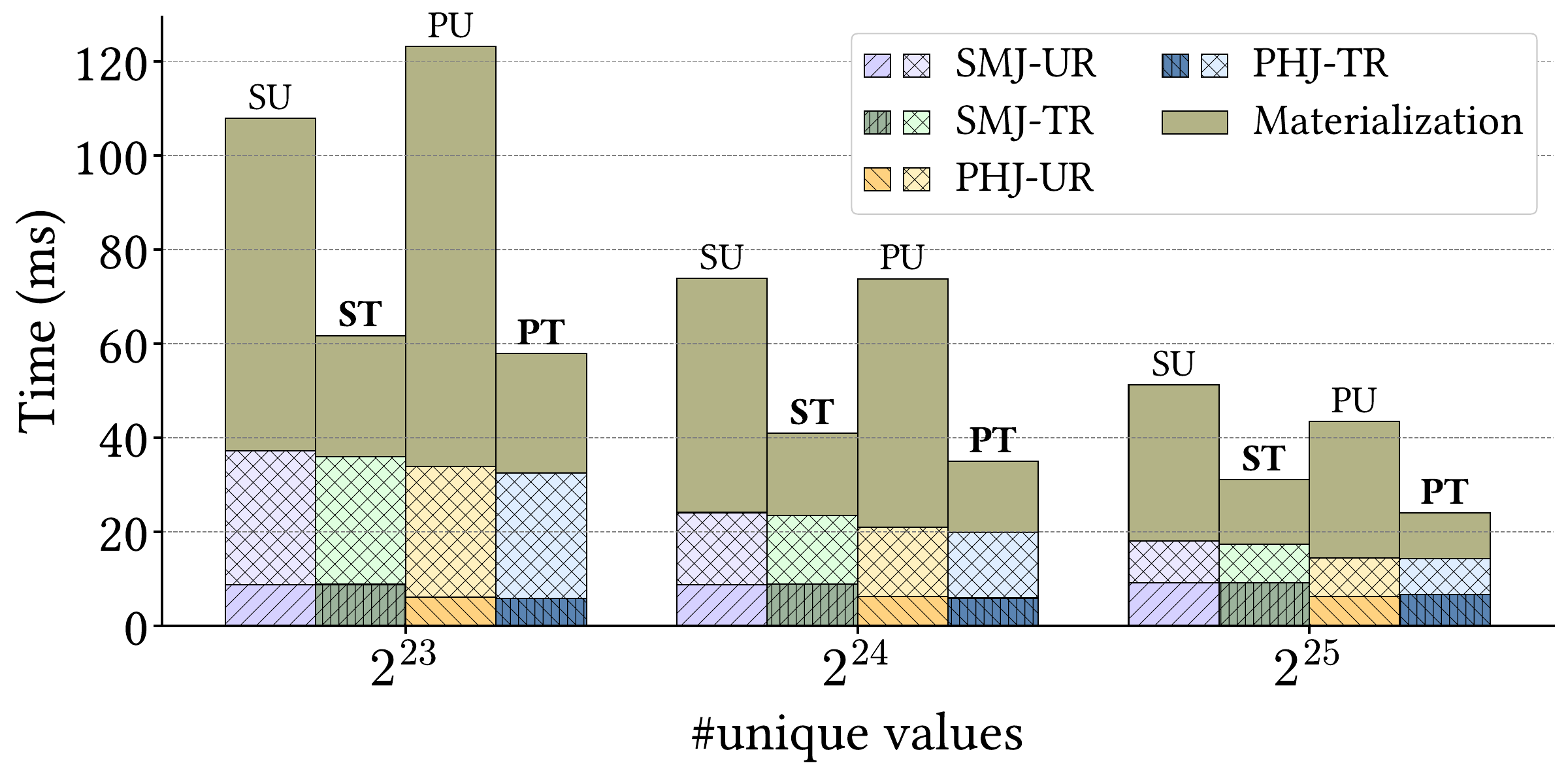}
        \caption{\color{ETHc} Many-to-many joins.}%
        \label{fig:fkfk}
    \end{minipage}\hfill
\end{figure*}

{\color{ETHc} So far, we have evaluated the algorithms with key-foreign-key joins and highlighted the advantage of our proposed GFTR technique. Next, we will demonstrate the effectiveness of GFTR in many-to-many joins. In this case, the output size can be larger than either input, resulting in high materialization costs. We generate the join attributes of $R$ and $S$ from a set of unique keys $K$ uniformly. $R$ and $S$ carry two columns of payload data, respectively. We set $|R|=|S|=2^{26}$ and change $|K|=2^{23}, 2^{24}, 2^{25}$. A smaller $K$ results in more matching tuples. Compared to one-to-many joins, many-to-many joins are more likely to produce more matches, hence a higher materialization cost. The result (see Figure~\ref{fig:fkfk}) demonstrates that the advantage of GFTR is also significant in many-to-many joins. When $|K|$ becomes smaller, GFTR-based implementations become more advantageous due to increasing output cardinality and higher materialization costs.}

\subsubsection{Summary}\label{sec:join-summary}
For narrow joins or wide joins with low match ratios, PHJ-* has better performance than SMJ-*, and PHJ-TR performs even better for skewed data due to more efficient partitioning. For wide joins, PHJ-TR and SMJ-UR are the best choices for dealing with skewed foreign keys. If the foreign keys are more uniformly distributed, and materialization time dominates the overall runtime, *-TR is better in most cases. With the presence of 8-byte values in keys and/or payloads, SMJ-TR experiences performance degradation and loses its competitiveness to *-UM, whereas PHJ-TR maintains its superiority. The reason that partition-based implementations dominate the algorithm choice in all scenarios is that partitioning is more efficient than sorting, but both transformations make the match-finding phase similarly efficient.

\subsection{Microbenchmarks for Group-By}
\subsubsection{Experiment Setup.} By default, we evaluate the performance of group-by on relation $R(k, r_1, r_2)$ with $k$ being the group key and $r_1$ and $r_2$ to be aggregated. We set $|R|=2^{28}$. The aggregate function considered here is maximum, which represents a large class of reducible aggregate functions like sum, product, and minimum. We evaluate both 4-byte and 8-byte group keys while fixing the non-key attributes to be 4-byte. Most of our evaluation works with keys sampled from a uniform distribution of $\{0,\dots,g-1\}$. We report the throughput measured in $|R|/\text{total time}$, and the total time includes necessary memory (de-)allocation, which are negligible. 
\subsubsection{Implementations.}
We evaluate seven implementations.
\begin{itemize}[wide, labelwidth=!, labelindent=0pt]
    \item [\textbf{cuDF}] Hash-based group-by from cuDF (v23.12) (Section~\ref{sec:hash-group-by}).
    \item[\textbf{SGB-UR}] Sort-based group-by with GFUR (Section~\ref{sort-groupby-gfur}).
    \item[\textbf{SGB-TR}] \textbf{(Ours)} Sort-based group-by with GFTR (Section~\ref{sort-groupby-gftr}).
    \item[\textbf{SGB-DE-TR}] \textbf{(Ours)} SGB-TR with dictionary encoding (Section~\ref{faster-sorting}).
    \item[\textbf{PGB-UR}] \textbf{(Ours)} Partition-based group by using GFUR (Section~\ref{sec:pgb}).
    \item[\textbf{PGB-TR}] \textbf{(Ours)} Partition-based group by using GFTR (Section~\ref{sec:pgb}).
    \item[\textbf{HGB}] \textbf{(Ours)} Optimized Hash-based group-by (Section~\ref{sec:opt-hash-groupby}). 
\end{itemize}

\subsubsection{Optimized hash-based group-by vs. cuDF}
\begin{figure}[t]
   \centering
    \includegraphics[width=0.48\linewidth]{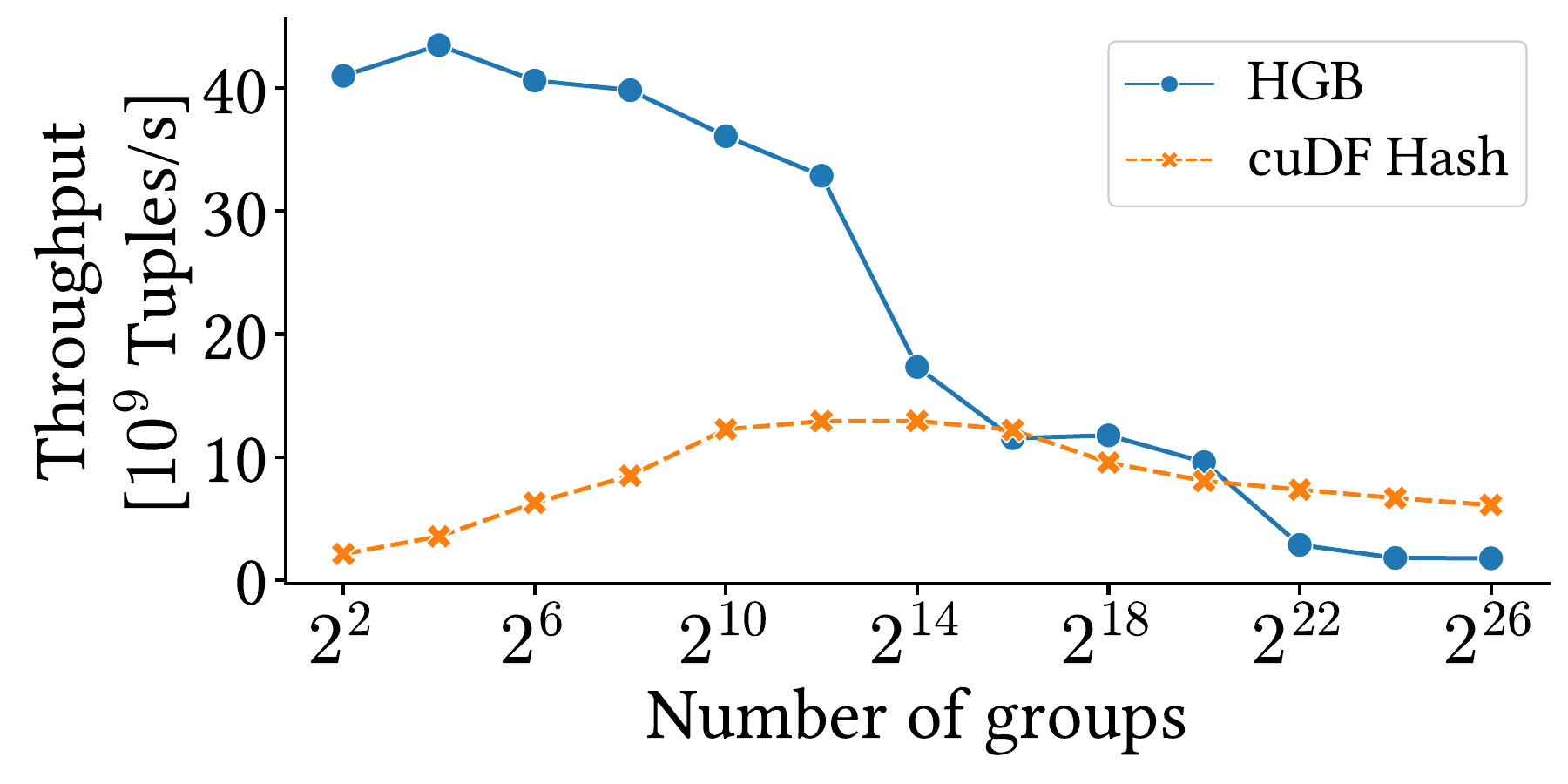}%
    \caption{HGB vs. cuDF.}%
    \label{fig:hgbvscudf}
\end{figure}

Figure~\ref{fig:hgbvscudf} shows that our optimized hash-based group-by (HGB) is significantly faster than cuDF for $g \le 2^{14}$. The performance benefit of HGB is more significant when $g$ is small since there is more contention for cuDF whereas HGB uses the shared memory to avoid contention when $g \le 2^{14}$. When $g \ge 2^{16}$, HGB has similar to worse performance compared to cuDF because HGB has more random accesses when computing the group membership.

Although cuDF is more competitive for extremely high $g$s, we will show later that partition-based and sort-based are better choices for this scenario. In other words, our HGB is much more efficient than cuDF for workloads where the hash-based group-by is preferred. We will only include our HGB for the rest of the evaluation to avoid bloating the analysis and figures.

\subsubsection{Partition-based Group-by}
Both implemented with GFTR, as shown in Figure~\ref{fig:groupbyall4b}, the partition-based group-by (PGB-TR) performs substantially better than the sort-based (SGB-TR and SGB-DE-TR) for all $g > 2^4$ (by up to 68\%). When $g \le 2^4$, PGB-TR suffers from having only a few thread blocks processing a large amount of data, a situation that can be alleviated by the measures mentioned in Section~\ref{sec:pgb}. However, when $g > 2^4$, each thread block is assigned a smaller amount of data, and the advantage of partitioning over sorting becomes more manifest.
Compared with HGB, PGB-TR is advantageous when $g > 2^{18}$. 

\subsubsection{Effectiveness of GFTR}
\begin{figure*}[t]
    \begin{minipage}[t]{.48\linewidth}
        \centering
        \includegraphics[width=1\linewidth]{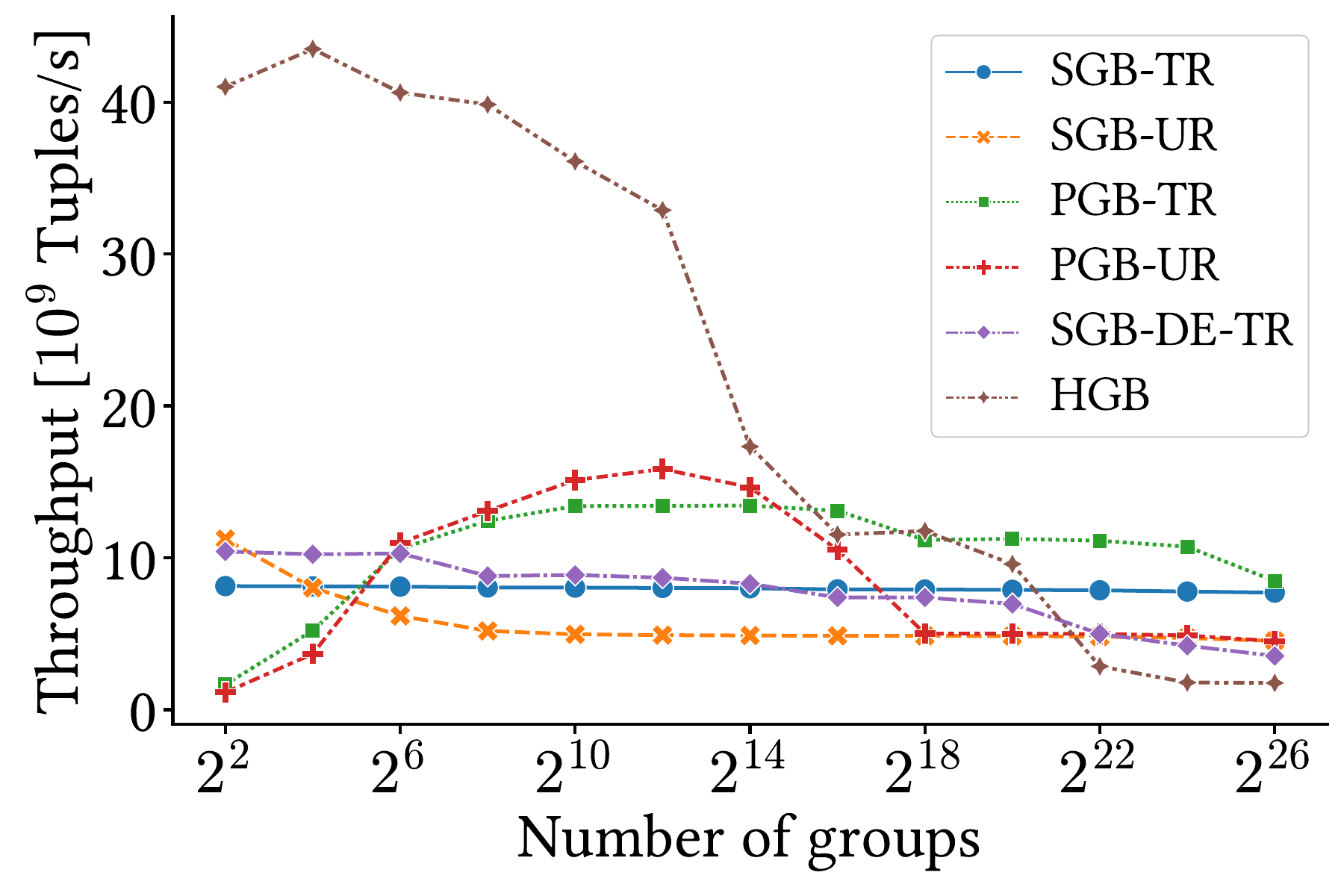}%
        \caption{Group-by with 4-byte keys.}%
        \label{fig:groupbyall4b}
    \end{minipage}\hfill
    \begin{minipage}[t]{.48\linewidth}
        \centering
        \includegraphics[width=1\linewidth]{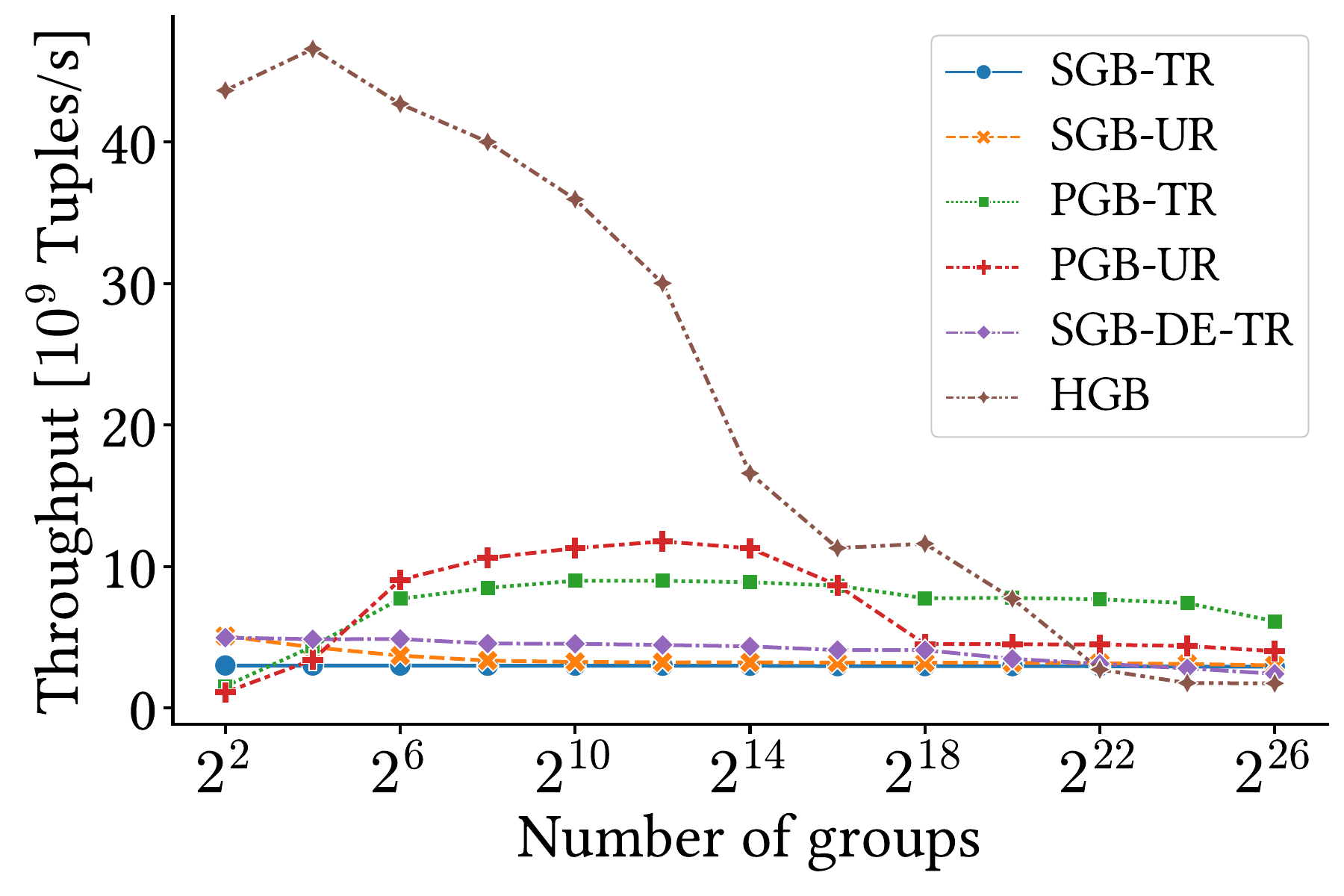}%
        \caption{Group-by with 8-byte keys.}%
        \label{fig:groupbyall8b}
    \end{minipage}
\end{figure*}

Figure~\ref{fig:groupbyall4b} shows the performance comparison between GFUR and GFTR versions of sort-based and partition-based group-by, respectively. SGB-TR is consistently faster than SGB-UR for $g > 2^4$ by up to 1.7x. For $g \le 2^4$, SGB-UR performs better due to fewer random accesses.
{\color{ETHc} Figure~\ref{fig:sgb-model1} shows the measured and model-predicted (see Section~\ref{sort-groupby-gftr}) time differences between SGB-UR and SGB-TR across different numbers of groups. The model predicts the turning point (i.e., the time difference is 0) to be around $g=2^3$, very close to the actual turning point $g=2^4$. When the number of groups is lower than $2^8$, our model overestimates the advantage of SGB-TR over SGB-UR. The potential reason could be the caching effect of the L2 cache, which our model does not capture. When the number of groups is greater than $2^8$, the model-predicted difference largely agrees with the measured difference.}

\begin{figure}[H]
    \centering
    \resizebox{0.45\linewidth}{!}{\begin{tikzpicture}
        \begin{axis}[
            title={},
            xlabel={Number of groups},
            ylabel={Time difference (ms)},
            xmin=0, xmax=28,
            ymin=-12, ymax=28,
            xtick={2,4,6,8,10,12,14,16,18,20,22,24,26},
            xticklabels={$2^{2}$,,$2^{6}$,,$2^{10}$,,$2^{14}$,,$2^{18}$,,$2^{22}$,,$2^{26}$},
            ytick={-10, -5,0,5,10,15,20,25},
            legend pos=south east,
            ymajorgrids=true,
            axis x line=bottom,        
            axis y line=left,          
            height=4cm,
            width=10cm
        ]
        
        \definecolor{mygreen}{rgb}{0.0, 0.6, 0.3}
        
        \addplot[
            color=blue,
            dashed,
            mark=*,
            ]
            coordinates {
(2,-3.45254605787781)(4,13.1196750199357)(6,24.1678224051447)(8,24.1678224051447)(10,24.1678224051447)(12,24.1678224051447)(14,24.1678224051447)(16,24.1678224051447)(18,24.1678224051447)(20,24.1678224051447)(22,24.1678224051447)(24,24.1678224051447)(26,24.1678224051447)
            };
        
        \addlegendentry{\texttt{Estimated}}
        
        \addplot[
            color=blue,
            mark=*,
            ]
             coordinates {
(2,-9.091643)(4,0.407219000000005)(6,10.332778)(8,18.388222)(10,20.694983)(12,21.19173)(14,21.357382)(16,21.258955)(18,21.292969)(20,21.342109)(22,21.657203)(24,22.32109)(26,24.482475)
            };
        \addlegendentry{\texttt{Real}}
        \end{axis}
\end{tikzpicture}}%
    \caption{\color{ETHc} Time(SGB-UR) - Time(SGB-TR)}
    \label{fig:sgb-model1}
\end{figure}

PGB-UR and PGB-TR perform similarly for $g \le 2^{14}$. 
They experience significant performance drops from $g = 2^{16}$ to $g = 2^{18}$ because each partition starts to contain more than one group. However, the performance of PGB-UR drops much more rapidly due to the growing number of random accesses, making PGB-TR up to 2.2x faster than PGB-UR.

The results demonstrate that the GFTR optimization is effective for both sort-based and partition-based algorithms. For the sort-based, GFTR takes effect for a very small number of groups whereas for the partition-based, it is only helpful for a large number of groups but the speedup is substantial.

\subsubsection{Effectiveness of dictionary encoding}
Figure~\ref{fig:groupbyall4b} shows SGB-DE-TR improves upon SGB-TR by up to 28\% when $g < 2^{16}$. This observation agrees with our model prediction in Table~\ref{tab:de-opt}. Noticeably, the performance of SGB-DE-TR drops abruptly from $g=2^6$ to $g=2^8$. This is because when the keys are encoded with $2^8$ alphabets, it needs one more pass of radix partitions for sorting (See Section~\ref{sec:gpu-primitives}). For 8-byte group keys (shown in Figure~\ref{fig:groupbyall8b}), the benefit of using the dictionary encoding to accelerate sorting is more pronounced -- bringing up to 65\% speedup and yielding better performance up to $2^{22}$ groups, which also agrees with our model.

{\color{ETHc} To demonstrate the predictability of our model in Section~\ref{faster-sorting}, we show the measured and model-predicted time differences between SGB-TR and SGB-DE-TR across different numbers of groups. For 4-byte keys, the model prediction almost completely agrees with the measurement. For 8-byte keys, the model prediction agrees with the measurement in trends while consistently underestimating the advantage of SGB-DE-TR over SGB-TR. }

\begin{figure}[H]
    \centering
    \resizebox{0.45\linewidth}{!}{\begin{tikzpicture}
        \begin{axis}[
            title={},
            xlabel={Number of groups},
            ylabel={Time difference (ms)},
            xmin=0, xmax=28,
            ymin=-45, ymax=40,
            xtick={2,4,6,8,10,12,14,16,18,20,22,24,26},
            xticklabels={$2^{2}$,,$2^{6}$,,$2^{10}$,,$2^{14}$,,$2^{18}$,,$2^{22}$,,$2^{26}$},
            ytick={-40,-30,-20,-10,0,10,20,30},
            ymajorgrids=true,
            axis x line=bottom,        
            axis y line=left,          
            height=4cm,
            width=10cm,
            legend style={font=\small,legend pos=south west,legend columns=2}
        ]
        
        \definecolor{mygreen}{rgb}{0.0, 0.6, 0.3}
        
        \addplot[
            color=blue,
            dashed,
            mark=*,
            ]
            coordinates {
(2,6.44475262422287)(4,5.06373406730969)(6,4.14305444630219)(8,4.14305197684881)(10,1.3810052527331)(12,1.3809657414791)(14,1.380807696463)(16,-3.2243431031083)(18,-5.9922802966774)(20,-6.0158816857449)(22,-6.110287242015)(24,-33.8781081929261)(26,-42.9841984205788)
            };
        
        \addlegendentry{\texttt{Est.(4-byte)}}
        
        \addplot[
            color=blue,
            mark=*,
            ]
             coordinates {
(2,7.221222)(4,6.801485)(6,7.039808)(8,2.854302)(10,3.117129)(12,2.585676)(14,1.210624)(16,-2.43161)(18,-2.416513)(20,-4.52091799999999)(22,-19.580623)(24,-29.165413)(26,-40.922887)
            };
        \addlegendentry{\texttt{Real(4-byte)}}

        \addplot[
            color=red,
            dashed,
            mark=*,
            ]
            coordinates {
(2,28.3108776437299)(4,26.239349792926)(6,24.8583302996784)(8,24.858326348553)(10,22.0962736977492)(12,22.0962104797428)(14,22.0959576077171)(16,15.1881681903537)(18,12.4170279511254)(20,12.3806143794212)(22,-27.0377513157556)(24,-29.4329551434083)(26,-41.7758073003215)
            };
        
        \addlegendentry{\texttt{Est.(8-byte)}}
        
        \addplot[
            color=red,
            mark=*,
            ]
             coordinates {
(2,35.355648)(4,34.213143)(6,34.439169)(8,30.880676)(10,30.507194)(12,29.556549)(14,28.315927)(16,25.078784)(18,24.943336)(20,13.444933)(22,4.997956)(24,-4.125602)(26,-19.122083)
            };
        \addlegendentry{\texttt{Real(8-byte)}}
        \end{axis}
\end{tikzpicture}}%
    \caption{\color{ETHc} Time(SGB-TR) - Time(SGB-DE-TR)}
    \label{fig:sgb-model2}
\end{figure}

\subsubsection{Data Skew}
\begin{figure}[t]
    \centering
    \begin{subfigure}[b]{.48\linewidth}
        \centering
        \includegraphics[width=\linewidth]{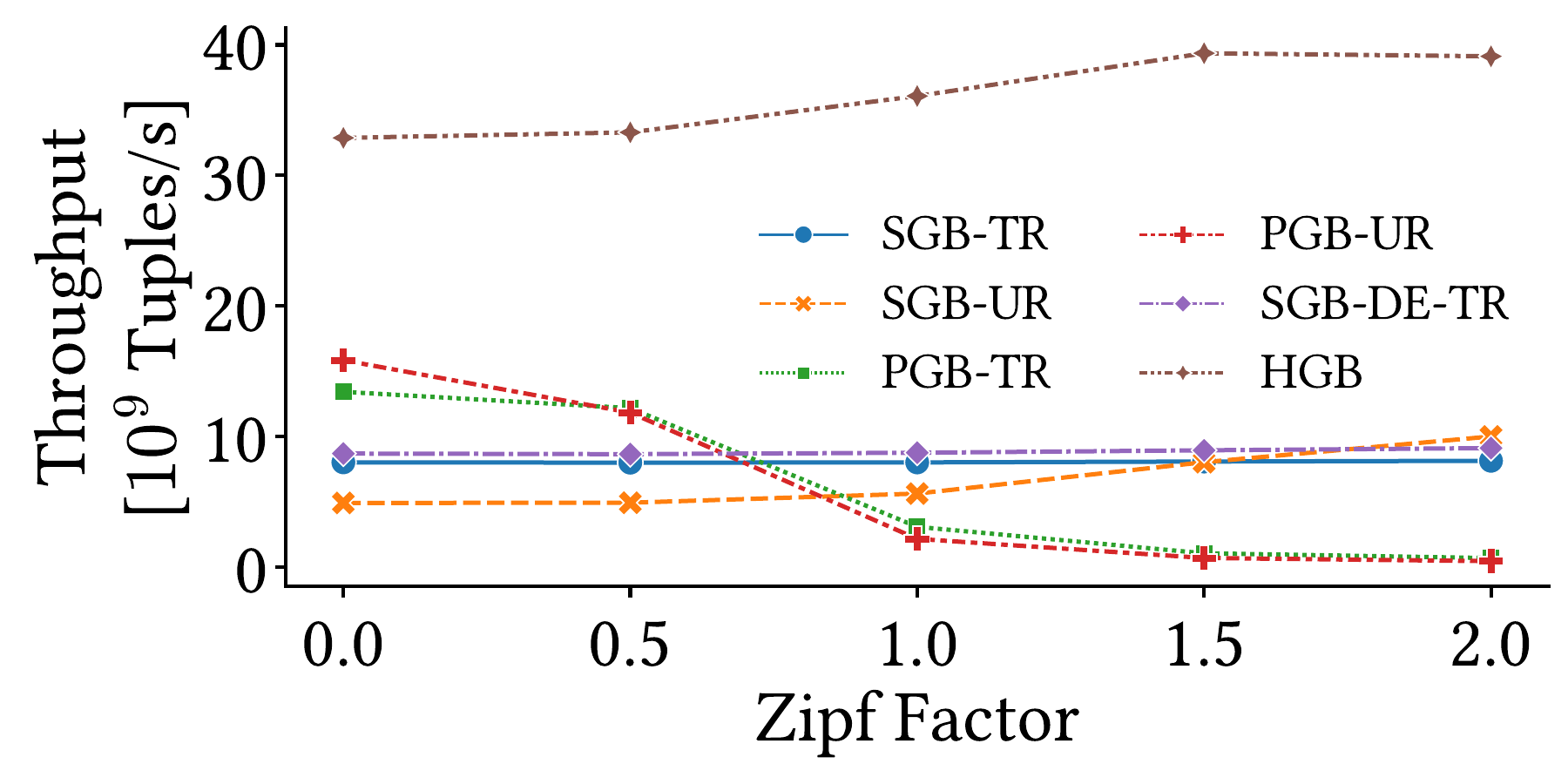}
        \caption{$g=2^{12}$}
    \end{subfigure}
    \begin{subfigure}[b]{.48\linewidth}
        \centering
        \includegraphics[width=\linewidth]{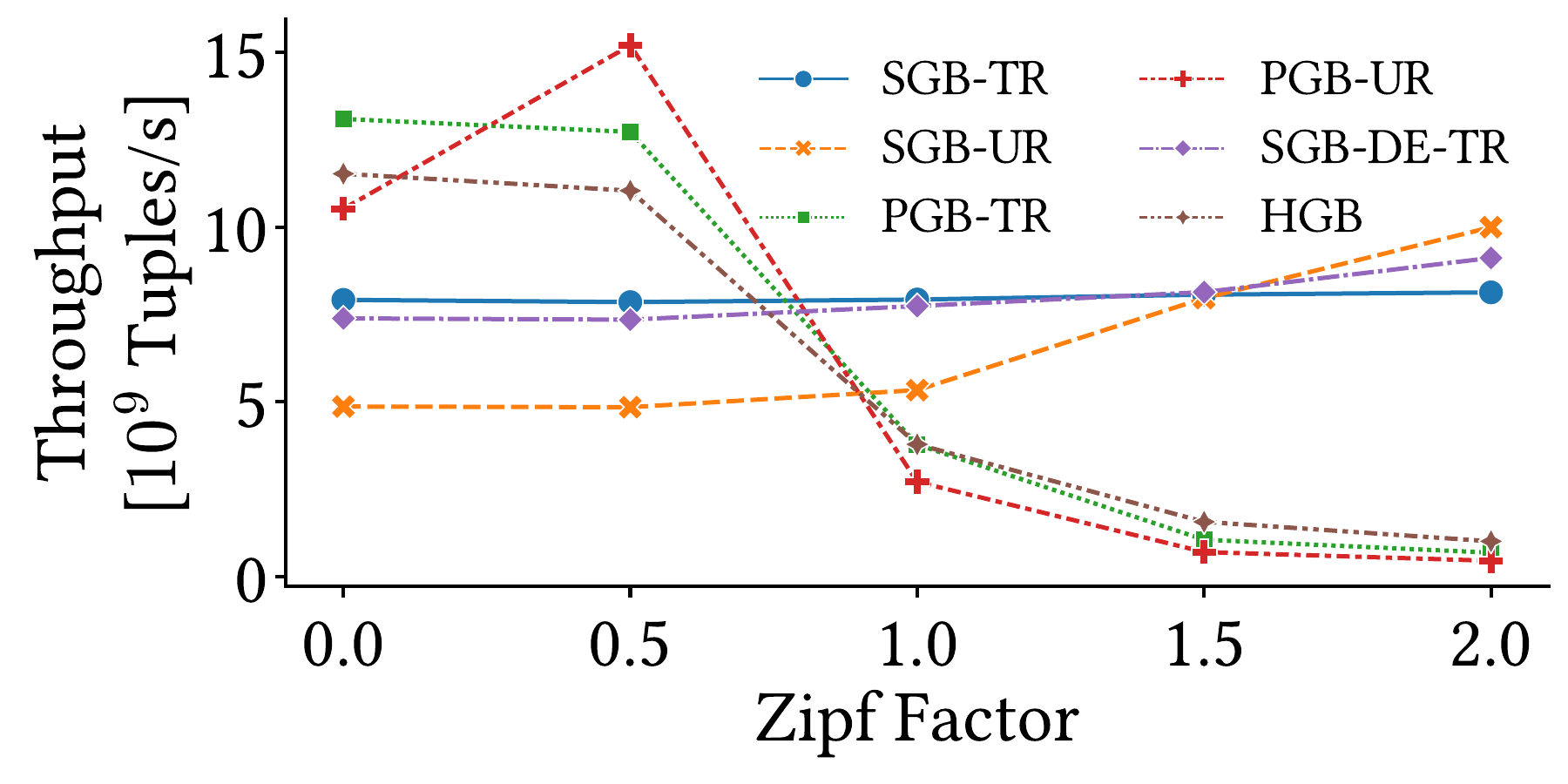}
        \caption{$g=2^{16}$}
    \end{subfigure}
    \caption{Effect of data skew on group-by.}
    \label{fig:groupbyskew}
\end{figure}

So far, we have assumed each group has the same number of tuples. To study the effect of data skew, we generate the group keys from the Zipfian distribution and adjust the level of skew by varying the Zipfian factor $z$, a common practice in previous work~\cite{Balkesen13-partitioned-hash-join,sioulas19-partitioned-radix-join, Bandle21-nonpartition-vs-partition}. The result is shown in Figure~\ref{fig:groupbyskew} (left). 
We fix $g=2^{12}$ in this experiment. 
As $z$ increases, sort-based algorithms can maintain at least the same performance. 
SGB-UR performs even better with increased $z$ because random accesses decrease when the data distribution is more concentrated. 
HGB is also very robust to data skew when $g=2^{12}$.
For partition-based approaches, the performance decreases substantially with growing $z$ because some large partitions are stressing some thread blocks. 

It is worth mentioning that the effect of the zip factor is intertwined with that of the number of groups. For example, when $g=2^{16}$ (Figure~\ref{fig:groupbyskew} right), the hash-based group-by loses the performance for high skew. This is because it cannot utilize the shared memory for $g=2^{16}$ and high skew causes contention of global memory accesses. 

\subsubsection{Data Types}
The length of the group key can substantially impact the group-by performance since it affects the cost of sorting and partitioning.
Figure~\ref{fig:groupbyall8b} shows the group-by performance evaluated with 8-byte keys. GFTR-based algorithms only perform better than the GFUR-based ones for a larger number of groups than 4-byte group keys. This is because sorting and partitioning become more expensive whereas unclustered \textsc{gather}s almost have the same performance for 4-byte and 8-byte. SGB-UR has better performance than SGB-TR for all $g$s evaluated; on the other hand, the optimized SGB-DE-TR outperforms SGB-UR for most cases by up to 39\%. Although PGB-UR performs better than PGB-TR for $g<2^{16}$, PGB-TR maintains its superiority again for $g > 2^{16}$ 
HGB has very significant advantages over other approaches when $g \le 2^{18}$ and quickly loses its dominance afterward.

\subsubsection{Summary}\label{sec:groupby-summary}
\begin{figure}[b]
    \centering
    \resizebox{0.65\linewidth}{!}{\begin{tikzpicture}[
    level 1/.style={sibling distance=45mm},
    level 2/.style={sibling distance=35mm},
    level 3/.style={sibling distance=45mm},
    level 4/.style={sibling distance=40mm},
    edge from parent/.style={draw, thick},
    edge from parent path={(\tikzparentnode.south) -- (\tikzchildnode.north)},
    every node/.style={align=center}
]

\node {Must use sort? Or group-by and\\order-by use identical keys?}
    child { node {Conditions in Table~\ref{tab:de-opt} met?}
        child { node {SGB-DE-TR}
        edge from parent node[left=1mm] {yes} }
        child { node {SGB-TR}
        edge from parent node[right=1mm] {no} }
    edge from parent node[left=1mm] {yes} }
    child { node {Can HGB use shared\\memory for aggregation?}
        child { node {HGB}
        edge from parent node[left=1mm] {yes} }
        child { node {High data skew?}
            child { node {Must use sort. Revisit the\\decision tree from root.}
            edge from parent node[left=1mm] {yes} }
            child { node {PGB-TR}
            edge from parent node[right=1mm] {no} }
        edge from parent node[right=1mm] {no} }
    edge from parent node[right=5mm] {no} };
\end{tikzpicture}}%
    \caption{Decision tree for group-by.}
    \label{fig:groupby-dt}
\end{figure}

Our experimental evaluation of group-by algorithms reveals their strengths and weaknesses and validates the accuracy of our performance models. Based on these insights, we obtain the decision tree shown in Figure~\ref{fig:groupby-dt} to predict the best algorithm for each scenario. When the aggregation function can only be processed by the sort-based group-by (e.g., unique) or the following order-by sorts data by the group keys, GFTR-optimized sort-based group-by algorithms are preferred. Choosing between SGB-DE-TR and SGB-TR is dictated by the conditions specified in Table~\ref{tab:de-opt}. For other cases, HGB should be selected whenever it is possible to aggregate using the shared memory. If not possible (i.e., $g$ is large), we can turn to the PGB-TR if the data is not highly skewed ($z < 1$); otherwise a sort-based approach should be used.

\section{Related Work}
\textbf{GPU-based join.}
Rui et al.~\cite{Rui17-fastequijoin} designed partitioned radix joins and sort-merge joins for GPUs by utilizing new hardware features, such as atomics, larger register files, and shared memory. They focused on the speedup of the GPU-based implementations relative to the CPU-based without comparing these two GPU-based algorithms across various workloads. 
Sioulas et al.~\cite{sioulas19-partitioned-radix-join} proposed the bucket-chain-based partitioning algorithm to support their partitioned hash join, which has been thoroughly discussed in Section~\ref{sec:unopt-phj}. Paul et al.~\cite{Paul20-revisit-gpujoin} revisited multiple hash joins on the GPUs and concluded that the partitioned hash join proposed by Sioulas et al.~\cite{sioulas19-partitioned-radix-join} outperforms all other existing hash join implementations. Our newly proposed partitioned hash join using the radix-sort-based partitioning and GFTR achieves the same partitioning efficiency while fixing the fragmentation and non-determinism issues and greatly reducing the materialization cost.
{\color{ETHc} Shanbhag et al.~\cite{Shanbhag20-crystal} described and evaluated a GPU-based non-partitioned hash join implemented using their tile-based processing framework. However, their limitation is that the implementation cannot handle hash collisions. Therefore, to make a fair comparison, we chose cuDF as our non-partitioned hash join baseline.}
Cai et al. proposed a skew-conscious GPU-based partitioned hash join algorithm that can detect overly skewed partitions and use different algorithms for normal and skewed partitions. Their target scenario is mainly the foreign-foreign-key join with data skew in both relations. Doraiswamy et al.~\cite{Doraiswamy23-graphicsjoin} leverage the graphics pipeline architecture instead of CUDA programming to implement a hardware-agnostic join but without considering the materialization cost.
He et al.~\cite{He22-tensor-runtime} implemented the sort- and hash-based joins using tensor operations. Hu et al.~\cite{Hu22-tcudb} converted joins to matrix multiplications, which can then be computed efficiently using the Tensor Core Units. Sun et al.~\cite{sun23-mmjoin} showed that both tensor-based joins have worse performance than cuDF and the partitioned hash join from~\cite{sioulas19-partitioned-radix-join}. A variety of other join implementations have been proposed and studied for older GPUs~\cite{he08-relationaljoin}, multi-way joins~\cite{Lai21-multiwayjoins}, CPU-GPU co-processing~\cite{he13-coprocessing, sioulas19-partitioned-radix-join}, out-of-memory cases~\cite{Rui17-fastequijoin,sioulas19-partitioned-radix-join,Kaldewey12-revisitjoin}, multi-GPUs~\cite{Rui20-multigpujoin,Paul21-multigpujoin,thostrup-gpu-rdma-join}, fast interconnects~\cite{Lutz22-tritonjoin,Lutz20-nonpartitionedjoin}, or as part of a GPU-based query engines~\cite{Bress14-cogadb,He09-gdb,Shanbhag20-crystal,cudf}.

\noindent\textbf{GPU-based group-by.}
Karnagel et al.~\cite{Tomas15-groupby} described and compared hash-based and sort-based group-by and explored how they react to the number of groups and various configurations (e.g., kernel launch parameters). They concluded that the hash-based group-by is almost always better than the sort-based and provided ways to tune the hash-based implementation. Our work indirectly compares with them because their implementation resembles the cuDF one. Since we assume the group cardinality $g$ to be unknown, we do not investigate building a hash table in the shared memory, which is mentioned by Karnagel et al. Tom\'e et al. explored the design space of group-by in the context of CPU-GPU co-processing with a special focus on accelerating TPC-H Q1. Rosenfeld et al.~\cite{rosenfeld-hash-groupby} evaluated how different parallelization strategies and thread configurations can affect the performance of hash-based group-by. They proposed an algorithm to tune these relevant parameters during query execution to maximize the performance. Although Karnagel et al. and Rosenfeld et al. discussed fine-tuning the group-by according to the workload, the query optimization discussed in our work focuses on choosing \emph{different} algorithms/implementations for different workloads instead of tuning a single implementation. Kroviakov et al.~\cite{kroviakov-crossdevice} studied the group-by across the CPU and GPU based on fragment-based parallelism.

\section{Conclusion}
In this work, we demonstrate that random accesses are the major cost of existing GPU-based join and group-by implementations, and we propose a technique called GFTR, to remove almost all random accesses in the sort-based and partition-based implementations. When the join has more than one non-key column (i.e., wide joins) and a high match ratio, GFTR can accelerate the state-of-the-art partitioned hash join~\cite{sioulas19-partitioned-radix-join} by up to 2.3x and sort-merge join by 1.6x. The threshold match ratio can be estimated by our model described in Section~\ref{sec:opt-phj}. GFTR can also improve the sort-based and partition-based group-by when the number of groups exceeds an implementation-dependent threshold. This threshold can be estimated by the model discussed in Section~\ref{sort-groupby-gftr}. Besides match ratio and group cardinality, we also dive deep into how to choose between GFTR and GFUR depending on other workload characteristics, like data skew and data types. 

Besides optimizing for random accesses, we also explore techniques to improve the group-by. We redesign the hash-based group-by by utilizing the shared memory for aggregation without assuming the group cardinality upfront. To reduce the high sorting costs, we propose two solutions, designing a partition-based group-by and using dictionary encoding to compress group keys. The partition-based group-by performs the best when the group cardinality is very high. Compressing group keys reduces the sorting cost substantially especially when the group key is 8-byte long. 

We evaluate the implementations thoroughly with various workloads. 
The evaluation results offer valuable heuristics for the query optimizer to select the most efficient join and group-by implementations. For joins, we focus on the case where both relations are large. Our partitioned hash join implemented with GFTR delivers the best performance when the match ratio is moderate or higher ($\ge 25\%$), when the foreign key distribution is skewed, or when the keys are 8 bytes. Otherwise, a GFUR-based partitioned hash join implementation is preferred.
For group-by operations, the evaluation demonstrates that our hash-based group-by outperforms alternatives across a wide range of group cardinalities ($\le 2^{18}$). In other cases when the group key distribution is uniform, the partition-based group-by offers better performance. For scenarios requiring sort-based algorithms—such as when handling complex aggregation functions—the dictionary encoding technique enhances sort-based group-by efficiency when the number of groups is relatively small. Additionally, the cost models developed in this work effectively characterize the conditions under which each implementation works the most efficiently.

In future work, we plan to extend our evaluation to cover more database operators. Additionally, we aim to perform similar performance analyses on AMD GPUs. For further integration, we intend to incorporate the operators and query optimizer into a full-fledged database system and assess their performance on TPC-H and TPC-DS queries.


\bibliographystyle{ACM-Reference-Format}
\bibliography{main}

\received{July 2024}
\received[revised]{September 2024}
\received[accepted]{November 2024}

\end{document}